\documentclass[lettersize,journal]{IEEEtran}
\usepackage{amsmath,amsfonts}
\usepackage{array}
\usepackage{textcomp}
\usepackage{stfloats}
\usepackage{url}
\usepackage{verbatim}
\usepackage{graphicx}
\usepackage{cite}
\usepackage{latexsym}

\usepackage{microtype}

\usepackage{booktabs}
\usepackage{amsmath}
\usepackage{algorithm}
\usepackage{algpseudocode}
\usepackage{multirow}
\usepackage{mathtools}
\usepackage{subfigure}
\usepackage{graphicx}
\usepackage{balance}
\usepackage{color}
\usepackage{xcolor}
\usepackage{enumitem}
\usepackage{xspace}
\usepackage{setspace}

\renewcommand{\vec}[1]{\ensuremath{\mathbf{#1}}}
\newcommand{\stitle}[1]{\vspace{1mm} \noindent {\bf #1}}
\newcommand{\eg}{{\it e.g.}}

\newcommand{\ie}{{\it i.e.}}

\newcommand{\bL}{\ensuremath{\mathcal{L}}}
\newcommand{\bD}{\ensuremath{\mathcal{D}}}

\newcommand{\bS}{\ensuremath{\mathcal{S}}}
\newcommand{\bQ}{\ensuremath{\mathcal{Q}}}

\newcommand{\bG}{\ensuremath{\mathcal{G}}}
\newcommand{\bN}{\ensuremath{\mathcal{N}}}
\newcommand{\bE}{\ensuremath{\mathcal{E}}}

\newcommand{\model}{G2P2}

\usepackage{caption}

\begin{document}

\title{Prompt Tuning on Graph-augmented Low-resource Text Classification}


\author{Zhihao Wen and Yuan Fang~\IEEEmembership{Senior Member,~IEEE}
\thanks{
Zhihao Wen and Yuan Fang (corresponding author) are with School of Computing and Information Systems, Singapore Management University, Singapore, 188065.
}
\thanks{E-mail: zhwen.2019@phdcs.smu.edu.sg and yfang@smu.edu.sg}
}



\markboth{IEEE Transactions on Knowledge and Data Engineering}%
{Shell \MakeLowercase{\textit{et al.}}: Bare Demo of IEEEtran.cls for Computer Society Journals}



\maketitle

\begin{abstract}
Text classification is a fundamental problem in information retrieval with many real-world applications, such as predicting the topics of online articles and the categories of e-commerce product descriptions.
However, low-resource text classification, with no or few labeled samples, presents a serious concern for supervised learning.
Meanwhile, many text data are inherently grounded on a network structure, such as a hyperlink/citation network for online articles, and a user-item purchase network for e-commerce products.
These graph structures capture rich semantic relationships, which can potentially augment low-resource text classification. 
In this paper, we propose a novel model called Graph-Grounded Pre-training and Prompting (\model) to address low-resource text classification in a two-pronged approach. During pre-training, we propose three graph interaction-based contrastive strategies to jointly pre-train a graph-text model; during downstream classification, we explore 
handcrafted discrete prompts and continuous prompt tuning 
for the jointly pre-trained model to achieve zero- and few-shot classification, respectively. 
Moreover, we explore the possibility of employing continuous prompt tuning for zero-shot inference. Specifically, we aim to generalize continuous prompts to unseen classes while leveraging a set of base classes. To this end, we extend  \model\ into \model$^*$, hinging on a new architecture of conditional prompt tuning.
Extensive experiments on four real-world datasets demonstrate the strength of \model\ in zero- and few-shot low-resource text classification tasks, and illustrate the advantage of \model$^*$ in dealing with unseen classes.
\end{abstract}


\maketitle

\section{Introduction}\label{sec:intro}
Text classification is a fundamental research problem with many important applications in information retrieval. For example, predicting the topics of online articles can help readers easily search and navigate within the website or portal \cite{mccallum2000automating}, and classifying the category of e-commerce product descriptions enables businesses to structure their inventory efficiently and improve users' search experience \cite{xu2019open}.
Recent advances in natural language processing (NLP) have achieved remarkable success for text classification, especially when there are large-scale and high-quality labeled data. However, data labeling is often costly and time-consuming, making low-resource classification, in which no or few labeled samples are available, an appealing alternative. 


To address low-resource text classification, one approach is to utilize pre-trained language models (PLM) \cite{kenton2019bert, radford2018improving}, many of which are based on the transformer architecture \cite{vaswani2017attention} due to its powerful ability of encoding texts. 
A PLM can be adapted to different tasks by \emph{fine-tuning} the model parameters to task-specific objectives. While the ``pre-train, fine-tune'' paradigm requires fewer labeled data than traditional supervised learning, it suffers from two drawbacks. Firstly, state-of-the-art PLMs typically have huge model size, \eg,  GPT-3 has 175 billion parameters \cite{brown2020language}, which makes fine-tuning prohibitively expensive \cite{li2021prefix}. Secondly, fine-tuning still needs a reasonable amount of  labeled data due to the gap between pre-training and fine-tuning objectives, and thus struggles with low-resource scenarios including zero- and few-shot classification.  

To overcome the problem of pre-training and fine-tuning, \emph{prompting} \cite{brown2020language} has been proposed. It uses a natural language instruction or ``prompt'' to give a hint of the downstream task, whilst freezing the parameters of a large PLM. In other words, no fine-tuning or additional training is required at all for a new task. 
However, discrete natural language prompts can be difficult to design and may result in suboptimal performance compared to fine-tuning \cite{lester2021power}. More recently,
\emph{prompt tuning} \cite{liu2021gpt, lester2021power} formulates a continuous prompt as a learnable embedding, which is optimized without updating the PLM. 

Meanwhile, text data frequently rely on network structures, such as hyperlink/citation networks for online articles or user-item interaction graphs for e-commerce products. These graph structures expose valuable relationships between content or descriptions, aiding low-resource text classification. While current PLMs and prompting do not leverage these relationships, graph neural networks (GNNs) \cite{wu2020comprehensive} excel in processing graph data. GNNs typically leverage a message-passing architecture, which allows for the integration of node features and topological structures, resulting in impressive performance on graphs.
Nevertheless, traditional end-to-end training of GNNs heavily relies on abundant task-specific labels, which motivates self-supervised GNNs \cite{wu2021self} using well-designed pretext tasks derived from a label-free graph in a contrastive \cite{velickovic2019deep} or generative \cite{hu2019strategies,hu2020gpt} manner.
However, the treatment of text features in GNNs remains rudimentary. A simple bag-of-words representation \cite{yang2016revisiting} or aggregation of shallow word embeddings \cite{mikolov2013efficient} is fed into GNNs as the ``initial message'', which are further propagated along graph structures. 
Hence, the modeling of texts is coarse-grained, unable to fully capture the subtle semantic differences and similarities within texts. 


\stitle{Present work: G2P2.}
To overcome the limitations of existing text- and graph-based solutions, we propose a novel approach called Graph-Grounded Pre-training and Prompting (\textbf{\model}). \model\ attempts to address the following two open questions.



Firstly, \emph{how do we capture fine-grained textual semantics, while leveraging graph structure information jointly?}
A na\"ive approach is to use a language model to generate features from raw texts as input, and then train a GNN. However, in this way the texts and graph are only loosely coupled, lacking an explicit pairing to complement each other.  
In this paper, we propose graph-grounded contrastive pre-training, to maximize the alignment between text and graph representations based on three types of graph interaction, namely, text-node, text-summary, and node-summary interactions.

Secondly, \emph{how do we augment low-resource text classification given a jointly pre-trained graph-text model?} 
Instead of following the traditional fine-tuning paradigm, we try to ``prompt'' our jointly pre-trained graph-text model, from which the most relevant structural and semantic information can be located to improve low-resource classification. Without the need to update a large pre-trained model, prompting is also more efficient than fine-tuning. Specifically, we employ discrete prompts in zero-shot classification and continuous prompts in few-shot settings. 
While discrete prompts are manually crafted in the absence of class labels, continuous prompts can be automatically learned from the few-shot labels through a prompt-tuning process. 
On one hand, discrete prompts (\textbf{\model+d}) often require intensive human labor, and the performance of tuned prompts can be uneven across tasks. On the other hand,  continuous prompt leverages the inherent differentiability of neural networks to tune the optimal prompts, requiring minimal human input. 
Prompt-tuning is both data- and computation-efficient owing to the much fewer parameters in a continuous prompt than in the pre-trained model.
Furthermore, considering the graph structures between texts, we propose a context-based initialization for prompt tuning, which could provide a more informative starting point. 

\stitle{Extension: \model$^*$.}
While continuous prompt performs better and requires less human effort than discrete prompts, tuning continuous prompts still need some labeled data and thus would not work for zero-shot inference.
Thus, let us consider an alternative zero-shot classification setting with both \emph{base classes} and \emph{unseen classes}: Each base class includes a small labeled set of instances for selecting discrete prompts (\model+d) or tuning continuous prompts (\model), while each unseen class has no labeled set.

\begin{figure}[t]
   \includegraphics[width=1\linewidth]{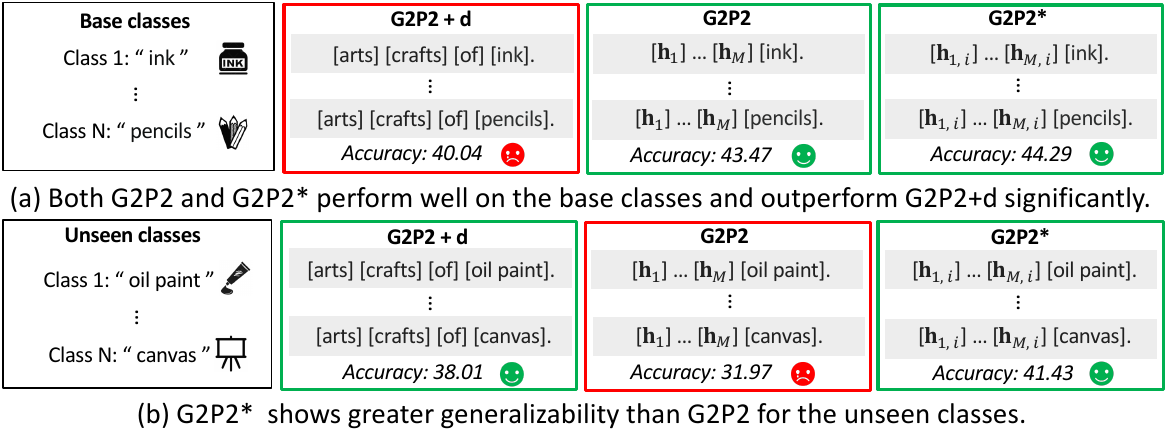}
   \vspace{-2mm}
	\caption{From \model\ to \model$^*$: Learning generalizable continuous prompts. Illustrations are based on the Amazon Art dataset (see Sect.~\ref{sec:expt:setup}).}
	\label{fig:motivation}
        \vspace{-4mm}
\end{figure}

Under this new zero-shot setting, continuous prompts can be tuned on the base classes, but the tuned prompts tend to overfit to the base classes and fall short in extending their reach to broader, unseen classes. 
For example, in Fig.~\ref{fig:motivation}(a), while \model\ demonstrates strong performance on the base classes like ``ink" or ``pencils", its performance declines sharply when applying the tuned prompts on the unseen classes like ``oil paint" or ``canvas". 
The significant drop in performance is primarily attributed to the potential class shifts often present even within the same domain, which hinders  \model's ability to generalize effectively to unseen classes. Therefore, an important research question emerges: \emph{How to generalize continuous prompts from base classes to unseen classes?}

To enhance the generalizability of \model, we propose an novel alternative \model$^*$, which substitute the vanilla prompt tuning with conditional prompt tuning on graphs. \model$^*$ expands upon \model\ by integrating a lightweight neural network that generates an input-conditioned prompt token for each node. In contrast to the static prompts of \model\ that are overly specific to the base classes, the conditional prompts of \model$^*$ are dynamically responsive to individual nodes, thereby providing greater robustness against class shifts in unseen classes \cite{zhou2022conditional}.
As illustrated in Fig.~\ref{fig:motivation}, \model$^*$ not only aligns well with the base classes, but also sustains a robust performance on the unseen classes, thereby showcasing its enhanced generalization capabilities.
Meanwhile, we note that handcrafted discrete prompts (\model+d) show a reasonable level of generalizability. However, they demand extensive manual work and their overall performance on both base and unseen classes is comparatively modest.

Note that \model$^*$ is not exclusively designed for zero-shot scenarios but serves as an enhancement to G2P2, aiming to improve its handling of zero-shot conditions while preserving its performance in few-shot situations. Essentially, our objective is to maintain strong performance across both base and new class scenarios. 

\stitle{Contributions.}
We make the following contributions. 
(1) This is the first attempt to pre-train text and graph encoders jointly for low-resource text classification.
(2) We propose a novel model called Graph-Grounded Pre-training and Prompting (\model) with three graph interaction-based constrastive strategies in pre-training, as well as discrete and continuous prompts for downstream zero- and few-shot classification, respectively.
(3) We extend beyond \model\ towards handling unseen classes, and propose \model$^*$, a conditional prompt tuning approach to generalize the continuous prompts to wider unseen classes while leveraging the base classes. 
(4) We conduct extensive experiments on four real-world datasets to demonstrate the strength of \model\ in zero- and few-shot text classification, and the ability of \model$^*$ to generalize to unseen classes.

A preliminary version of this paper has been published in SIGIR'23 \cite{wen2023augmenting}. We provide a summary of the major additions in this version. (1) \textbf{Problem}: In Sect.~\ref{sec:intro}, we introduced a new problem setting in which zero-shot inference on unseen classes is performed while leveraging a set of base classes. Furthermore, we restructured the introduction to emphasize the motivation, challenges and insights of generalizing prompt tuning to broader unseen classes.
(2) \textbf{Methodology}: 
To tackle the limited generalizability of \model, we extended \model\ and introduced a novel conditional prompt tuning approach, named \model$^*$, in Sect.~\ref{sec:conditional}. The conditional prompt tuning can be achieved by incorporating an additional lightweight neural network trained on the base classes. This network aims to generate an input-dependent token for each input node, which is then fused with learnable global prompt vectors to acquire a customized prompt for each node. 
(3) \textbf{Experiments}: 
In Sect.~\ref{sec:expt:conditional}, we conducted additional experiments to evaluate the ability 
of \model$^*$ to generalize from base classes to unseen classes. The results demonstrate its excellent generalization capabilities, outperforming \model\ by 3.5--29.6\% for within-domain generalization and 8.8--29.8\% for cross-domain generalization.

\begin{figure*}[t]
  \centering
  \includegraphics[width=1\linewidth]{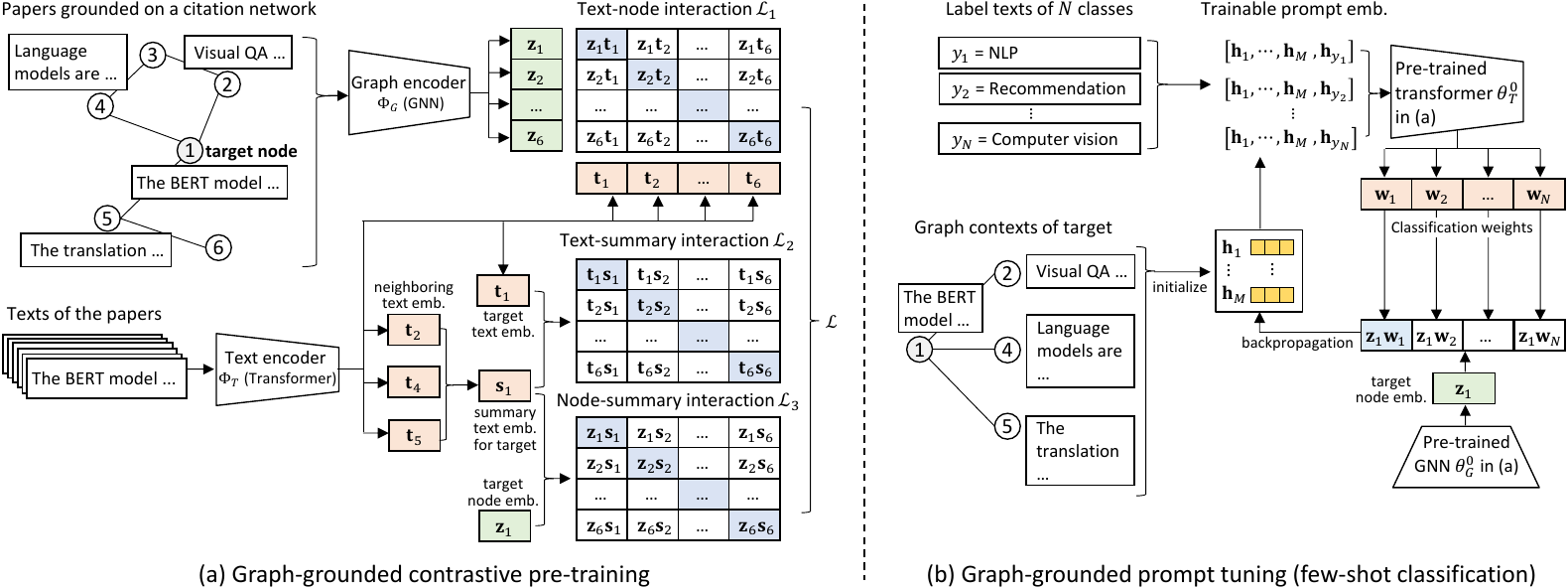}
  \vspace{-2mm}
  \caption{Overall framework of \model. (a) During pre-training, it jointly trains a text and a graph encoder through three contrastive strategies. (b) During testing, it performs prompt-assisted zero- or few-shot classification. Note that part (b) only shows continuous prompt tuning for few-shot classification, while discrete prompts  for zero-shot inference and conditional prompt tuning for generalization to wider unseen classes are presented separately in Figs.~\ref{fig:zero-shot} and \ref{fig:meta net}, respectively.}
   \vspace{-4mm}
  \label{fig:framework}
\end{figure*}


\section{Related Work}\label{sec:relatedwork}

\stitle{Graph neural networks.}
Inspired by the success of CNN in computer vision, GNNs have emerged to handle non-Euclidean relational data \cite{wu2020comprehensive}, ranging from early semi-supervised models such as GCN \cite{KipfW17}, GAT \cite{velivckovic2018graph} and GIN \cite{XuHLJ19}, to the more recent self-supervised pre-training paradigm \cite{velickovic2019deep, hu2019strategies, hu2020gpt, lu2021learning}.
Besides their widespread success on graph tasks, they have also been leveraged to improve text-based tasks through knowledge graphs \cite{cao2021dekr} and heterogeneous graphs \cite{linmei2019heterogeneous}, or multi-modality learning \cite{liu2021mm}. 
However, these approaches either employ coarse-grained text treatment, or have decoupled graph and text encoders without fully exploiting the intrinsic relationship between them. Although GLEM \cite{zhao2023learning} integrating both the text and graph structure information with large language models and GNNs, it is not a good low-resource learner.

\stitle{Language pre-training and prompting.}
Pre-trained language models \cite{han2021pre} have become the most popular backbone in NLP. While earlier PLMs such as GPT \cite{radford2018improving}, BERT \cite{kenton2019bert}, XLNet \cite{yang2019xlnet} and RoBERTa \cite{liu2019roberta} still have affordable model size, recent introductions such as T5 \cite{raffel2020exploring} and GPT-3 \cite{brown2020language} produce massive models with billions of parameters.
To avoid the high fine-tuning cost on these large models, prompting \cite{liu2021pre} starts to receive more attention.
A prompt is a special template to pad the task input, with a goal of extracting useful knowledge from PLMs to flexibly adapt to downstream tasks.
Fueled by the success of GPT-3, numerous prompting methods including discrete natural language prompt \cite{shin2020autoprompt, schick2021exploiting, gao2021making} and continuous prompt \cite{li2021prefix, lester2021power, liu2021gpt, qin2021learning,  zhong2021factual} have emerged. The strength of prompting has been validated in a wide range of NLP applications, including text classification \cite{hu2022knowledgeable, min2022noisy,sun2021nsp, zhang2021aspect, han2021ptr}, machine translation \cite{tan2022msp} and relation extraction \cite{chen2022knowprompt, sainz2021label}. More recently, prompting has also been applied to GNNs for graph-centric tasks such as node classification \cite{sun2022gppt, tan2023virtual} and graph classification  \cite{fang2022prompt, liu2023graphprompt}, but they cannot utilize fine-grained text information. 

\stitle{Zero- or few-shot paradigms.}
Broadly speaking, our setting is also related to other  learning paradigms. For example, in semi-supervised learning \cite{MiyatoDG17, xie2020unsupervised, chen2020mixtext}, each class may only have a few examples, but all classes must be seen in training and they cannot handle any novel class during testing.
Meta-learning \cite{finn2017model, yu2018diverse, han2018fewrel, bansal2020self, bao2020few, zhou2019meta, wang2020graph, wen2021meta, wen2023generalizing} 
is another popular paradigm that supports few-shot learning. However, large-scale labeled data are still required in a so-called ``meta-training'' phase, to support the few-shot learning of novel classes during ``meta-testing''.
In contrast, we only need label-free data for pre-training, without requiring any meta-training phase that would consume large-scale labeled data. 
Separately, there also exists joint consideration of image and text data using a contrastive pre-training strategy for zero- or few-shot classification
 \cite{radford2021learning, zhou2022learning, zhou2022conditional}. In our work, graph data are significantly different from images, which provide various types of interaction between texts. On graphs, zero-shot node classification has also been done \cite{wang2021zero}. It relies heavily on the availability of Wikipedia pages or other side information to generate class prototype embeddings. However, it is very labor-intensive to find and curate the right side information, especially when there are a large number of classes and/or novel classes emerge frequently.

\section{Preliminaries} 

In this section, we introduce relevant concepts and our low-resource classification settings. 

\stitle{Graph-grounded text corpus.} Consider a set of documents $\bD$, which is grounded on a graph $\bG=(\bD, \bE, \vec{X})$ such that each document $d_i\in\bD$ is a node $v_i$ in the graph. The documents are linked via edges in $\bE$, which are formed based on the application (\eg, if each document represents an article, the edges could be citations between articles). Each node $v_i$ is also associate with a feature vector $\vec{x}_i$, given by the input feature matrix $\vec{X}$. Finally, each document/node\footnote{We will use ``node'' and ``document'' interchangeably given their one-one correspondence in our context.} has a class label (\eg, the topic of the article). 


\stitle{Low-resource classification.}
A low-resource task consists of a support set $\bS$ and a query set $\bQ$. The support set $\bS$ contains $N$ classes, and each class has $K$ labeled examples where $K$ is a small number (\eg, 1 or 5), known as $N$-way $K$-shot classification. The query set $\bQ$ contains one or more unlabeled instances belonging to the $N$ classes in the support set. Our goal is to classify the instances in the query set based on the labeled examples in the support set.
Unlike episodic few-shot meta-learning \cite{finn2017model} which has both training tasks and testing tasks, we only have testing tasks; in the training stage, we perform self-supervised pre-training on label-free data only. 
As a special case, tasks with $K=0$ are known as zero-shot classification, which means there is no labeled example at all and we can only rely on class metadata (\eg, class label text).



\section{Proposed Approach: \model} 
In this section, we introduce our novel model \model\ for low-resource text classification.

\subsection{Overview of \model}

As shown in Fig.~\ref{fig:framework}, our model \model\ consists of two stages: (a) graph-grounded constrastive pre-training, and (b) prompt-tuning for low-resource classification. 
 
During pre-training as shown in Fig.~\ref{fig:framework}(a),  we learn a dual-modal embedding space by jointly training a text encoder and graph encoder in a self-supervised fashion, since a document also exists as a node on the graph. More specifically, we use a transformer-based text encoder and a GNN-based graph encoder. The transformer takes the text on each node (\ie, document) as the input, and outputs a text embedding vector $\vec{t}_{i}$ for node $v_i$. On the other hand, the GNN takes the graph and node features as input, and generates a node embedding vector $\vec{z}_i$ for node $v_i$. Subsequently, in the dual-modal embedding space, we align the text and graph representations on the same or related nodes through three contrastive strategies based on different types of interaction on the graph. 

In downstream testing, 
we use handcrafted discrete prompts (Fig.~\ref{fig:zero-shot})  together with the label text for zero-shot classification. For few-shot classification, we use continuous prompts to pad the label text (Fig.~\ref{fig:framework}(b)). 



\subsection{Graph-grounded contrastive pre-training}

As shown in Fig.~\ref{fig:framework}(a), the graph-grounded pre-training learns a dual-modal embedding space by jointly training a text encoder and a graph encoder, based on three types of interaction on the underlying graph.

\stitle{Dual encoders.} 
The text encoder is a transformer \cite{vaswani2017attention}, which we denote $\Phi_T$. Given a document $d_i$, the text encoder
outputs the $d$-dimensional embedding vector of $d_i$, denoted $\vec{t}_i\in \mathbb{R}^d$:
\begin{align}
\label{eq:text emb}
    \vec{t}_i = \Phi_T(d_i;\theta_T),
\end{align}
where $\theta_T$ represents the parameter set of the transformer.
Correspondingly, let $\vec{T} \in \mathbb{R}^{|\bD|\times d}$ represents the text embedding matrix for all documents.

Meanwhile, a document $d_i$ is also a node $v_i$ in the graph. We choose a classic GNN called graph convolutional network (GCN) \cite{KipfW17} as the graph encoder, denoted $\Phi_Z$. It similarly outputs an embedding vector $\vec{z}_i\in \mathbb{R}^d$ for a given node $v_i$:
\begin{align}
\label{eq:node emb}
    \vec{z}_i= \Phi_Z(v_i;\theta_G), 
\end{align}
where $\theta_G$ represents the parameter set of the GCN.
Likewise, let $\vec{Z} \in \mathbb{R}^{|\bD|\times d}$ represents the graph embedding matrix for all nodes.

\stitle{Text-node interaction.}
\label{sec:model:node level}
Our graph-grounded texts naturally implies a bijection between nodes and texts, where each document $d_i$ corresponds to the node $v_i$ in the graph.
Inspired by the pairing of image and its caption text \cite{radford2021learning} and the mapping of content and node sequences \cite{liu2018content}, we design a pre-training strategy to predict which text document matches which node in the graph.


Specifically, given $n$ documents and the corresponding $n$ nodes, there are $n^2$ possible document-node pairs $\{(d_i,v_j)\mid i,j =1,\ldots,n \}$.
Among them, only $n$ pairs with $i=j$ are true matching, whereas the remaining $n^2-n$ pairs are false matching. 
%
%
As our first contrastive strategy, we exploit the bijective interaction between texts and nodes on the graph, to maximize the cosine similarity of the $n$ matching pairs, while minimizing the cosine similarity of the $n^{2}-n$ unmatching pairs. To compute the cosine similarity for the $n^2$ pairs, we first perform a row-wise L2 normalization on embedding matrices $\vec{T}$ and $\vec{Z}$ to obtain $\tilde{\vec{T}}$ and $\tilde{\vec{Z}}$, respectively. We   then compute a node-text similarity matrix $\vec{\Lambda}_1 \in \mathbb{R}^{n\times n}$ to capture the pairwise cosine similarity, as follows.
%
\begin{align}
\label{eq:sm1}
    \vec{\Lambda}_{1}= \left( \tilde{\vec{Z}} \tilde{\vec{T}}^{\top} \right) \cdot \exp(\tau), 
\end{align}
where $\tau\in \mathbb{R}$ is a trainable temperature parameter to scale the similarity values \cite{radford2021learning}. 

\textsc{\textbf{Remark.}} Although $\vec{\Lambda}_{1}\in\mathbb{R}^{n\times n}$ is a dense matrix, it is constructed in a batch-wise manner for practical implementation. That is, $n$ is not the total number of documents but the relatively small batch size, and thus the overhead is negligible. $\vec{\Lambda}_{2}$ and $\vec{\Lambda}_{3}$ will be introduced later following the same treatment. 

To formulate the contrastive loss based on the text-node bijective interaction, we adapt the \emph{multi-class N-pair loss} \cite{sohn2016improved,zhang2020contrastive}, by considering both the row-wise and column-wise cross entropy loss w.r.t.~the row or column index. For instance, the $i$-th row of $\vec{\Lambda}_1$ represents the similarity scores between node $v_i$ and every document, in which the row index $i$ indicates the ground truth document $d_i$ that matches $v_i$.
\begin{align}
\label{eq:node level}
    \bL_1 = \textstyle\frac{1}{2}\left(\textsc{CE}(\vec{\Lambda}_1, \vec{y}) + \textsc{CE}(\vec{\Lambda}_1^{\top}, \vec{y}) \right),
\end{align}
where $\vec{y}=(1,2,\ldots,n)^\top$ is the label vector for contrastive training, and \textsc{CE} denotes the cross entropy loss applied to the input matrix $\vec{\Lambda}_1$ or $\vec{\Lambda}_1^\top$ in a row-wise manner.

\stitle{Text-summary interaction.}
\label{sec:model:TT}
Apart from the bijective text-node interaction, we further exploit  higher-order interactions on the graph. 
In particular, each document has a set of neighboring documents defined by graph topology. The neighboring documents can be understood as a summary of the target document given the semantic relatedness between them. For example, on a e-commerce network, the products purchased by a user naturally portray a summary of the user and vice versa. 
%
Without loss of generality, we employ a simple mean pooling to generate the summary embedding $\vec{s}_i\in \mathbb{R}^d$ as follows.
\begin{align}
\label{eq:context emb}
    \vec{s}_{i}=\textstyle\frac{1}{|\bN_i|} \sum_{j \in \bN_{i}} \vec{t}_{j}.
\end{align}
For efficiency, we only sample a fixed number of neighboring documents to generate the summary. 
Then, let $\vec{S} \in \mathbb{R}^{n\times d}$ denote the summary text embedding matrix for all documents.

Hence, as our second contrastive strategy, we seek to align the text embedding of each document and its corresponding summary text embedding, based on the text-summary interaction derived from graph neighborhood. In other words, we maximize the cosine similarity of the $n$ matching pairs of document and its neighborhood-based summary, while minimizing the cosine similarity of the $n^{2}-n$ unmatching pairs. 
Specifically, we first follow Eq.~\eqref{eq:sm1} to construct a text-summary similarity matrix  $\vec{\Lambda}_2 \in \mathbb{R}^{n\times n}$:
\begin{align}
\label{eq:sm2}
    \vec{\Lambda}_{2}= \left( \tilde{\vec{T}} \tilde{\vec{S}}^{\top} \right) \cdot \exp(\tau).
\end{align}
Subsequently, we apply the same contrastive loss following Eq.~\eqref{eq:node level}, as follows.
%
\begin{align}
\label{eq:TT}
    \bL_2 = \textstyle\frac{1}{2}\left(\textsc{CE}(\vec{\Lambda}_2, \vec{y}) + \textsc{CE}(\vec{\Lambda}_2^{\top}, \vec{y}) \right),
\end{align}

\stitle{Node-summary interaction.}
The neighborhood-based summary for document $d_i$ also serves as a semantic description of node $v_i$.
Mirroring the text-summary interaction, as our third contrastive strategy, we seek to align the node embedding and its neighborhood-based summary text embedding. We similarly compute a node-summary similarity matrix  $\vec{\Lambda}_3 \in \mathbb{R}^{n\times n}$, and formulate the corresponding contrastive loss $\bL_3$.
\begin{align}
    \vec{\Lambda}_{3}&= \left( \tilde{\vec{Z}} \tilde{\vec{S}}^{\top} \right) \cdot \exp(\tau),\label{eq:sm3}\\
    \bL_3 &= \textstyle\frac{1}{2}\left(\textsc{CE}(\vec{\Lambda}_3, \vec{y}) + \textsc{CE}(\vec{\Lambda}_3^{\top}, \vec{y}) \right).\label{eq:ZT}
\end{align}


\stitle{Overall pre-training objective.}
Finally, we integrate the three contrastive losses based on the text-node, text-summary and node-summary interactions. 
We obtain a pre-trained model $\theta^0=(\theta_{T}^0, \theta_{G}^0)$ consisting of the parameters of the dual encoders, given by 
\begin{align}
\label{eq:final loss}
    \theta^0 = \arg\min_{\theta_{T}, \theta_{G}} \bL_{1} + \lambda (\bL_{2} + \bL_{3}),
\end{align}
where $\lambda \in \mathbb{R}^+$ is a hyperparameter to balance the contribution from summary-based interactions. 

The pre-training procedure is outlined in Algorithm~\ref{alg:pre-train}, which has the following complexity per epoch. Let  $|\bD|$ be the number of documents, $\eta$ be the number of neighbors sampled to generate the summary embedding in Eq.~\eqref{eq:context emb}, and $\beta$ be the batch size. First, the cost of generating the three types of embeddings (lines 5--8) per epoch is  $O(|\bD|\eta)$, given that calculating the summary embedding needs go through $\eta$ neighbors.
Second, the cost of calculating the three similarity matrices in each batch is $O(\beta^2)$, and the total cost per epoch is $O\left(\frac{|\bD|}{\beta}\beta^2\right)=O(|\bD|{\beta})$ given $\frac{|\bD|}{\beta}$ batches in an epoch.
Thus, the overall complexity is $O(|\bD|(\eta+\beta))$, which is linear in the number of documents given that $\eta$ and $\beta$ are small constants. In our implementation, we set $\eta=3$ and $\beta=64$.
\begin{algorithm}[t]
\footnotesize
\caption{\textsc{Pre-training Procedure of \model}}
\label{alg:pre-train}
\begin{algorithmic}[1]
    \Require A graph-grounded text corpus $\bG=(\bD, \bE, \vec{X})$.
    \Ensure Pre-trained weights of text encoder $\theta^0_{T}$, graph encoder $\theta^0_{G}$.
    \State $\theta^0_{T},\theta^0_{G} \gets$ parameters initialization;
    \While{not converged}
        \State sample batches of documents from $\bD$;
        \For{each batch}
            \For{each node $v_{i}$/document $d_{i}$ in the batch} 
                \State compute $d_{i}$'s text embedding $\vec{t}_{i}$; \Comment{Eq.~\eqref{eq:text emb}}
                \State compute $v_{i}$'s node embedding $\vec{z}_{i}$;
                \Comment{Eq.~\eqref{eq:node emb}}
                \State compute $v_{i}$'s summary embedding $\vec{s}_{i}$;
                \Comment{Eq.~\eqref{eq:context emb}}
            \EndFor 
            \State compute similarity matrices $\vec{\Lambda}_{1},\vec{\Lambda}_{2},\vec{\Lambda}_{3}$;
            \Comment{Eqs.~\eqref{eq:sm1}, \eqref{eq:sm2}, \eqref{eq:sm3}}
            \State compute contrastive losses $\bL_{1}$, $\bL_{2}$, $\bL_{3}$;
            \Comment{Eqs.~\eqref{eq:node level},~\eqref{eq:TT},~\eqref{eq:ZT}}
            \State update the overall loss $\bL$;           
            \Comment{Eq.~\eqref{eq:final loss}}
            \State $\theta^0_{T},\theta^0_{G}\gets$ update via backpropagation  
        \EndFor 
    \EndWhile 
    \State \Return $\theta^0_{T},\theta^0_{G}$.
\end{algorithmic}
\end{algorithm}
\vspace{-4mm}

\subsection{Prompting joint graph-text model}\label{sec:model_prompt}
After pre-training our graph-text model, it is non-trivial to apply it to low-resource classification. To narrow the gap between pre-training and downstream tasks, the traditional ``pre-train, fine-tune'' paradigm typically introduces a new projection head for the downstream task, which will be fine-tuned together with the whole pre-trained model. However, under the low-resource setting, it is neither effective nor efficient to update the entire model with a huge number of parameters.
Without updating massive PLMs, prompting has recently emerged as a powerful alternative to fine-tuning in NLP  \cite{liu2021pre}, although prompting has not been explored for graph-text models with jointly pre-trained structural and textual  information. 

In this part, we elaborate zero-shot classification using handcrafted discrete prompts, as the absence of labeled data in zero-shot tasks cannot support directly learnable prompts. We further discuss automated continuous prompt tuning for few-shot classification.

\begin{figure}[t]
   \includegraphics[width=0.9\linewidth]{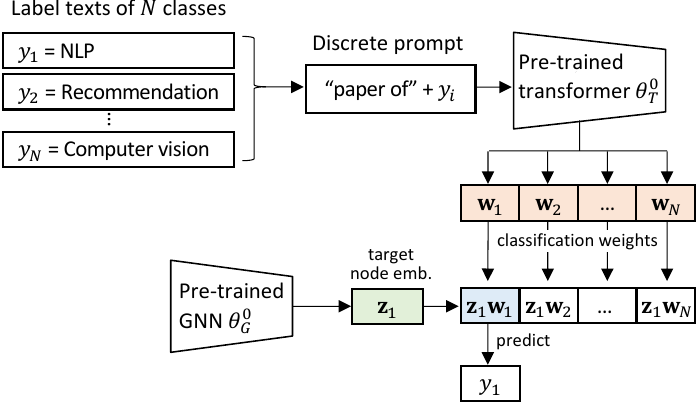}
   \vspace{-2mm}
	\caption{Schematic diagram for zero-shot classification. The pre-trained models $\theta_G^0$ and $\theta_T^0$ are obtained from Fig.~\ref{fig:framework}(a).}
	\label{fig:zero-shot}
        \vspace{-2mm}
\end{figure}

\stitle{Discrete prompts for zero-shot classification.}
In $N$-way zero-shot classification, out of $N$ classes, we predict the class which has the highest similarity to the given node. As illustrated by the diagram in Fig.~\ref{fig:zero-shot}, the classification weights can be generated by the text encoder based on the class label texts \cite{wang2018joint}, without requiring any labeled sample for the classification task. Specifically, the weight vector $\vec{w}_y$ for class $y\in\{1,2,\ldots,N\}$ is the output from the pre-trained text encoder, \ie,
\begin{align}
    \vec{w}_y=\phi_T(\text{``}\mathtt{prompt\ [CLASS]}\text{''}; \theta_T^0).\label{eq:discrete-prompt}
\end{align}
Here ``$\mathtt{prompt\ [CLASS]}$'' is a prompt template, where $\mathtt{[CLASS]}$ refers to the label text of the target class $y$ (\eg, ``$\mathtt{NLP}$'' for paper area classification), and $\mathtt{prompt}$ is a handcrafted sequence of natural language tokens to signal the relevance of the label text (\eg, ``$\mathtt{paper\ of\ NLP}$'' helps focus on the topic of the paper). In the simplest case, $\mathtt{prompt}$ can be an empty string so that we only rely on the label text. Note that discrete tokens are still converted to continuous word embeddings as input to the text encoder; for brevity we omit this step in Eq.~\eqref{eq:discrete-prompt}. 

Then, the class distribution given node representation $\vec{z}_i$ is predicted as 
\begin{align}
    p(y \mid \vec{z}_i)=\frac{\exp \left(\langle\vec{z}_i,\vec{w}_{y}\rangle\right)}{\sum_{y=1}^{N} \exp \left(\langle\vec{z}_i, \vec{w}_{y}\rangle\right)},\label{eq:softmax_clf}
\end{align}
where $\langle\cdot, \cdot\rangle$ is cosine similarity.

\stitle{Continuous prompts for few-shot classification.}
The problem with discrete prompts is that they are difficult to optimize, 
given that  PLMs are intrinsically continuous.
Substituting discrete natural language prompts with learnable continuous prompts, prompt tuning \cite{lester2021power,liu2021gpt,liu-etal-2022-p} can automate the optimization of prompts when some labeled data is available. 
Hence, in the few-shot setting, we explore prompt tuning to cue in the relevant structural and semantic information from our jointly pre-trained graph-text model.

Specicifally, instead of a sequence of discrete tokens, we take a sequence of 
continuous embeddings   $[\vec{h}_{1}, \cdots, \vec{h}_{M}, \vec{h}_\mathtt{CLASS}]$ as the prompt, 
where $M$ is a hyperparameter indicating the number of prompt tokens, each $\vec{h}_{m}$ ($m\le M$) is a trainable vector, 
and $\vec{h}_\mathtt{CLASS}$ is the word embedding sequence of the target class label. 
The continuous prompt is fed as input to the text encoder to generate the classification weights for each class $y$:
\begin{align}
\label{eq:prompt embeddings}
    \vec{w}_y=\phi_T([\vec{h}_{1}, \cdots, \vec{h}_{M}, \vec{h}_\mathtt{CLASS}]; \theta_T^0),
\end{align}
where each $\vec{h}_{m}$ ($m \le M$) has the same dimension as the input word embeddings to the text encoder. 
%

Using the same softmax layer in Eq.~\eqref{eq:softmax_clf}, 
we further update the continuous prompt embeddings using the labeled support set of the few-shot task by minimizing a cross entropy loss, whilst freezing the parameters of the dual encoders. This prompt tuning process is both data- and computation-efficient, given the small number of learnable parameters in the prompt. 
%



Furthermore, existing prompt tuning methods either initialize the prompt embeddings randomly \cite{lester2021power,liu-etal-2022-p} or using the word embeddings of handcrafted discrete prompts \cite{zhou2022learning}. While random initialization is non-informative and more prone to local optimum, it is still difficult to pick the right discrete prompts for initialization. 
 Therefore, we take the advantage of graph structures to initialize the prompt embeddings. 
 
 Specifically, given a node $v_i$, we define its \emph{graph contexts} as its neighbor set $\{v_j\mid j\in \bN_i\}$. Due to the underlying semantic relatedness, the graph contexts of the few-shot examples carry strong signals about the task, which can be exploited to improve the initialization.
For each document/node $v_i$ in the task support set, we sample $\eta$ nodes from its graph contexts. For $v_i$ itself and each context node sampled, we truncate its corresponding document to $M$ words, and convert it to a sequence of $M$ word embedding vectors, each having the same dimension as the vector $\vec{h}_m$ ($m\le M$) in our continuous prompt. Hence, for each support node, we would obtain $\eta+1$ such sequences; in an $N$-way $K$-shot task, there is 
a total of $NK(\eta+1)$ sequences.
We take the average of these embedding sequences to initialize the learnable prompt vectors $\vec{h}_1,\ldots,\vec{h}_M$, which is derived from graph contexts and thus could provide a more informative starting point than random initialization.

\section{Conditional prompt tuning: \model$^*$}
\label{sec:conditional}

In this section, we first review the limitations of \model\ in generalizing to unseen classes,
and propose \model$^*$ to overcome the limitation based on conditional prompt tuning.

\subsection{Limitation of \model}
The proposed continuous prompt tuning in Sect.~\ref{sec:model_prompt} aims to learn a collection of trainable vectors in each few-shot task. Compared to handcrafted discrete prompts, prompt tuning is automated and tends to be more robust than manual engineering. However, it still requires some labeled data, and thus has limited capability in dealing with
zero-shot inference, in which the test set involves the so-called ``unseen'' classes without any labeled instance. 
Due to class shift, directly using prompts learned from a set of ``base'' classes with labeled data do not generalize well to broader unseen classes without labeled data.

Specifically, consider a subset of classes in each dataset as \emph{base} classes and the rest as \emph{unseen} classes. In a base class, only a small number of labeled instances are available for selecting discrete prompts or tuning continuous prompts, reserving the majority for testing.  In contrast, all instances of the \emph{unseen} classes are used solely for testing, without revealing any labeled instance for prompt selection or tuning.
In this setting, we are essentially conducting zero-shot inferences on the unseen classes, using the prompts learned from the base classes.

As illustrated in Fig.~\ref{fig:motivation}, we can observe a critical limitation in \model. On one hand, in Fig.~\ref{fig:motivation}(a), \model\ can learn effective prompts to accurately identify base classes, such as ``ink'' and ``pencils''. On the other hand, in Fig.~\ref{fig:motivation}(b), when using the same prompts optimized for the base classes, the accuracy of \model\ markedly diminishes when facing novel, unseen classes like ``oil paint'' and ``canvas''. The significant decrease in performance occurs despite the consistent nature of the task, which involves the classification of item categories in the same Art domain. One potential reason is that the prompts learned by \model\ overfit to the base classes, resulting in poor generalization to unseen classes. The issue originates from the static prompt design, where the prompts, once learned, are tailored specifically to the base classes. Meanwhile, \model+d with handcrafted discrete prompts  demonstrates a comparatively higher level of generalization with only a small performance decrease on the unseen classes. However, handcrafted prompts require significant manual labor and can give inconsistent performance on different tasks or datasets. 


\begin{figure}[t]
   \centering
   \includegraphics[width=0.9\linewidth]{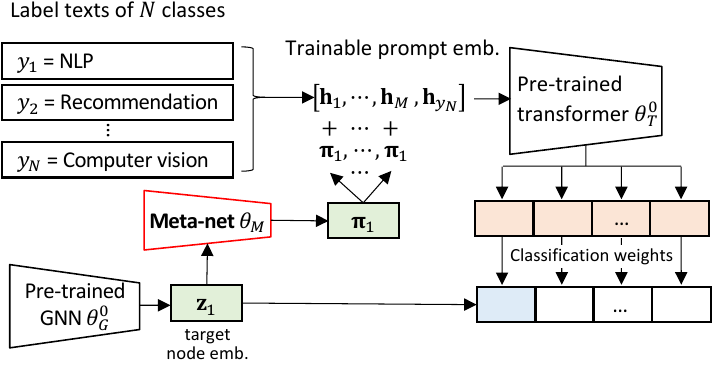}
	\caption{Schematic diagram for conditional prompt tuning in \model$^*$. The pre-trained models $\theta_G^0$ and $\theta_T^0$ are obtained from Fig.~\ref{fig:framework}(a). The classes are base class during tuning, and unseen classes during zero-shot inference.}
	\label{fig:meta net}
    \vspace{-2mm}
\end{figure}

\subsection{Proposed extension: \model$^*$}
To generalize prompt tuning to wider unseen classes, we explore conditional prompt tuning \cite{zhou2022conditional}. The core principle is to \emph{condition} a prompt on each individual input instance (\ie, each document/node in our context), rather than learning a static prompt tailed to a specific task or a fixed set of base classes. At the same time, to ensure parameter efficiency, we extend \model\ by adding a lean neural network to generate an input-conditional token (vector) for each node, which is then integrated with the learnable prompt vectors. We refer the conditional variant of our method as \model$^*$. 
Intuitively, the conditional token is analogous to node captioning or document summarization, which is equivalent to a  description for each node or document. Hence, conditional prompts are more generalizable: They are optimized to depict each instance and thus more resilient to class shift, rather than being confined to certain specific classes.

 Recall that in the unconditional version, we learn $M$ global prompt tokens that can be used with all nodes in a task. In contrast, in the conditional version, we need $M$ prompt tokens that are specific to each input node. A straightforward way is to increase the parameters, whereby we associate each node with $M$ learnable vectors directly, or train $M$ neural networks to generate $M$ distinct vectors for each node. In either design, the model size is substantially larger than the global prompt vectors in the original \model.
Inspired by the Meta-net architecture \cite{zhou2022conditional, ha2017hypernetworks}, we advocate for a more parameter-efficient design that has demonstrated impressive results in low-resource settings. Instead, we introduce a light-weight neural network, known as a Meta-net, on top of the $M$ global prompt vectors to generates one conditional token (vector) for each input node, which is subsequently fused with the global prompt vectors.

A schematic diagram of the conditional prompt tuning in \model$^*$ is sketched in Fig.~\ref{fig:meta net}.
Specifically, our Meta-net $\Phi_M$ employs a dual-layer multi-layer perceptron (MLP) with a bottleneck structure \cite{wu2018reducing}. 
Given a node $v_i$ with its embedding $\vec{z}_i$, the Meta-net generates a conditional token $\vec{\pi}_i$, \ie,
\begin{align}
\label{eq:meta net}
    \vec{\pi}_i= \Phi_M(\vec{z}_i;\theta_M).
\end{align}
Then, the $m$-th prompt vector for node $v_i$, denoted $\vec{h}_{m,i}$, is obtained by 
\begin{align}
\vec{h}_{m,i} = \vec{h}_{m} + \vec{\pi}_i,
\end{align}
where $\vec{h}_{m}$ is the $m$-th global prompt vector in the original \model. Note that the conditional token $\vec{\pi}_i$ should have the same dimension as the global prompt vector $\vec{h}_{m}$.
The conditional prompt for node $v_i$ is thus 
$[\vec{h}_{1,i}, \cdots, \vec{h}_{M,i}, \vec{h}_\mathtt{CLASS}]$,
where $\vec{h}_\mathtt{CLASS}$ is the word embedding sequence of the target class label identical to that used in the continuous prompt.
Subsequently, the conditional prompt is fed as input to the text encoder to output the classification weights, which means the weights are also conditioned on the input node. Specifically, the classification weight for node $v_i$ and class $y$ is
\begin{align}
\vec{w}_{y,i}=\phi_T([\vec{h}_{1,i}, \cdots, \vec{h}_{M,i}), \vec{h}_\mathtt{CLASS}]; \theta_T^0).
\end{align}
Thus, the prediction probability is computed as
\begin{align}
\label{eq:meta prob}
    p(y\mid\vec{z}_i)=\frac{\exp \left(\langle\vec{z}_i,\vec{w}_{y,i}\rangle\right)}{\sum_{y=1}^{N} \exp \left(\langle\vec{z}_i, \vec{w}_{y,i}\rangle\right)}.
\end{align}
The training process is similar to G2P2's prompt tuning: we train the M global prompt vectors and Meta-net by minimizing cross-entropy through few-shot node classification on the base classes.
Meta-net is just another way of prompt learning. It extends the single global prompt to node-specific prompts. Besides, meta-net is small, it is similar to other Parameter Efficient Fine-Tuning methods \cite{hu2021lora} that also use a small network to work with Large models.

\section{Experiments}
We conduct extensive experiments to evaluate \model\ and \model$^*$%
, with comparison to state-of-the-art baselines and detailed model analyses.

\subsection{Experimental setup}\label{sec:expt:setup}
\stitle{Datasets.}
Four public graph-grounded text corpora are used, as summarized in Tab.~\ref{table.datasets}.
The first dataset is \textbf{Cora}
\cite{mccallum2000automating}: known as the ``Cora Research Paper Classification'' dataset, it is a collection of research papers that are linked to each other through citations. 
    The abstract of a paper is deemed a text document. 
    The papers are classified into a topic hierarchy with 70 leaves. 
    Note that we are using a more comprehensive version of Cora, which is larger and has more classes than the version used elsewhere \cite{KipfW17}.
The other three datasets, namely, \textbf{Art}, \textbf{Industrial} and \textbf{Music Instruments (M.I.)}
, are all Amazon review collections \cite{ni2019justifying}, respectively from three broad areas, namely, arts, crafts and sewing (Art), industrial and scientific (Industrial), and musical instruments (M.I.). 
    Product descriptions and aggregated user reviews are treated as text, with reviews linking users to products and aiding in detailed classification into many specific subcategories. User reviews also enhance understanding of product-related text.
For all datasets, we employ the word2vec algorithm \cite{mikolov2013efficient} to obtain the 128-dimensional word embeddings of each word in the text documents. Then, for each node, we average the word embedding vectors of all the words in its document, and the averaged vector is used as the node's input features for the GNN-based methods. 

\begin{table}[t]
    \footnotesize
	\centering  
	\caption{Statistics of datasets.} 
        \vspace{-2mm}
	\label{table.datasets}  
	\begin{tabular}{l|rrrr}  
		\toprule  
		Dataset  & Cora    & Art       & Industrial & M.I. \\  \midrule
        \# Documents & 25,120  & 1,615,902 & 1,260,053  & 905,453        \\
        \# Links & 182,280 & 4,898,218 & 3,101,670  & 2,692,734     \\
        \# Avg.~doc length &141.26&54.23&52.15&84.66\\
        \# Avg.~node deg &7.26&3.03&2.46&2.97\\
        \# Total classes &70&3,347&2,462&1,191\\

		\bottomrule
	\end{tabular}
	\vspace{-4mm}
\end{table}

\stitle{Task construction.}
We perform zero- or few-shot text classification.
We adopt a \emph{5-way}  setting, \ie, we sample five classes from all the classes to construct a task. In each task, we construct a $K$-shot support set by further sampling $K$ examples from each class for $K \in \{0, 1,\ldots, 5\}$, and a validation set of the same size as the support set. The remaining examples form the query set.
Note that the support set is labeled and serve as the task training data, whereas the query set is unlabeled and used for evaluation. Note that in our experiment all the classes are used---it is only that each task involves 5 classes, and we have multiple tasks during testing to cover all the classes. This is a typical task setup \cite{finn2017model},  allowing for a comprehensive evaluation under different class combinations. The reported results are averaged over all the tasks on each dataset. 


\begin{table*}[tbp]
     \footnotesize 
	\centering 
	\caption{\emph{Five-shot} classification performance (percent) with 95\% confidence intervals.  
	} 
	\label{table:few-shot} 
    {
     \vspace{-2mm}
    \scriptsize In each column, the best result among all methods is \textbf{bolded} and the best among the baselines is \underline{underlined}. Improvement by \model\ is  calculated\\  relative to the best baseline.  $^{*}$ indicates that our model significantly outperforms the best baseline  based on the two-tailed $t$-test $(p<0.05)$.
    }
    \\[1.5mm] 
	\begin{tabular}{@{}c|cc|cc|cc|cc@{}}  
		\toprule
		  &\multicolumn{2}{c|}{Cora}&\multicolumn{2}{c|}{Art}&\multicolumn{2}{c|}{Industrial}&\multicolumn{2}{c}{M.I.}\\\cmidrule{2-9}
		  & Accuracy&Macro-F1&Accuracy&Macro-F1&Accuracy&Macro-F1&Accuracy&Macro-F1
		 \\\midrule
		 GCN &41.15$\pm$2.41 &34.50$\pm$2.23 &22.47$\pm$1.78 &15.45$\pm$1.14 
   &21.08$\pm$0.45 &15.23$\pm$0.29 &22.54$\pm$0.82 &16.26$\pm$0.72  \\
		 SAGE$_\text{sup}$  &41.42$\pm$2.90 &35.14$\pm$2.14 &22.60$\pm$0.56 &16.01$\pm$0.28 
   &20.74$\pm$0.91 &15.31$\pm$0.37 &22.14$\pm$0.80 &16.69$\pm$0.62  \\
        TextGCN &59.78$\pm$1.88 &55.85$\pm$1.50 &43.47$\pm$1.02 &32.20$\pm$1.30 
        &53.60$\pm$0.70 &45.97$\pm$0.49 &46.26$\pm$0.91 &38.75$\pm$0.78  \\
		 \midrule
		 GPT-GNN &76.72$\pm$2.02 &72.23$\pm$1.17 &65.15$\pm$1.37 &52.79$\pm$0.83 
   &62.13$\pm$0.65 &54.47$\pm$0.67 &67.97$\pm$2.49 &59.89$\pm$2.51  \\
		 DGI &\underline{78.42}$\pm$1.39 &\underline{74.58}$\pm$1.24 &65.41$\pm$0.86 &53.57$\pm$0.75 
   &52.29$\pm$0.66 &45.26$\pm$0.51 &68.06$\pm$0.73 &60.64$\pm$0.61  \\
		 SAGE$_\text{self}$ &77.59$\pm$1.71 &73.47$\pm$1.53 &76.13$\pm$0.94 &65.25$\pm$0.31 
   &71.87$\pm$0.61 &65.09$\pm$0.47 &\underline{77.70}$\pm$0.48 &\underline{70.87}$\pm$0.59  \\
		 \midrule
        BERT &37.86$\pm$5.31 &32.78$\pm$5.01 &46.39$\pm$1.05 &37.07$\pm$ 0.68
        &54.00$\pm$0.20 &47.57$\pm$0.50 &50.14$\pm$0.68 &42.96$\pm$1.02  \\
        BERT$^{*}$ &27.22$\pm$1.22 &23.34$\pm$1.11 &45.31$\pm$0.96 &36.28$\pm$0.71  
        &49.60$\pm$0.27 &43.36$\pm$0.27 &40.19$\pm$0.74 &33.69$\pm$0.72  \\
        RoBERTa &62.10$\pm$2.77 &57.21$\pm$2.51 &72.95$\pm$1.75 &62.25$\pm$1.33 
        &76.35$\pm$0.65 &70.49$\pm$0.59 &70.67$\pm$0.87 &63.50$\pm$1.11  \\
        RoBERTa$^{*}$ &67.42$\pm$4.35 &62.72$\pm$3.02 &74.47$\pm$1.00 &63.35$\pm$1.09 
        &77.08$\pm$1.02 &71.44$\pm$0.87 &74.61$\pm$1.08 &67.78$\pm$0.95  \\
		
		\midrule
        P-Tuning v2 &71.00$\pm$2.03 &66.76$\pm$1.95 &\underline{76.86}$\pm$0.59 &\underline{66.89}$\pm$1.14 
        &\underline{79.65}$\pm$0.38 &\underline{74.33}$\pm$0.37 &72.08$\pm$0.51 &65.44$\pm$0.63  \\
		
		\midrule
		\model-p &79.16$\pm$1.23 &74.99$\pm$1.35 &79.59$\pm$0.31 &68.26$\pm$0.43 
  &80.86$\pm$0.40 &74.44$\pm$0.29 &81.26$\pm$0.36 &74.82$\pm$0.45  \\
		\model\ &\textbf{80.08}$^{*}${}$\pm$1.33 &\textbf{75.91}$^{*}${}$\pm$1.39 
        &\textbf{81.03}$^{*}${}$\pm$0.43 &\textbf{69.86}$^{*}${}$\pm$0.67 
        &\textbf{82.46}$^{*}${}$\pm$0.29 &\textbf{76.36}$^{*}${}$\pm$0.25 
        &\textbf{82.77}$^{*}${}$\pm$0.32 &\textbf{76.48}$^{*}${}$\pm$0.52  \\
		(improv.) & (+2.12\%)&(+1.78\%) & (+5.43\%)&(+4.44\%) 
  & (+3.53\%)&(+2.7\%) & (+6.53\%)&(+7.92\%)\\
	\bottomrule
	\end{tabular}
	\vspace{-2mm}
\end{table*}

\stitle{Baselines for few-shot classification.}
We consider competitive baselines from four categories. 

(1) \emph{End-to-end GNNs}, which are graph neural networks trained in a supervised, end-to-end manner from random initialization, including: \textbf{GCN} \cite{KipfW17}; \textbf{SAGE$_\text{sup}$}, the supervised version of GraphSAGE \cite{hamilton2017inductive}; \textbf{TextGCN} \cite{yao2019graph}, a GCN-based model on a text graph constructed from word co-occurrence and document-word relations, which jointly learns the embeddings of both words and documents.

(2) \emph{Pre-trained/self-supervised GNNs}, which are pre-trained using pretext tasks without labeled data, followed by fine-tuning or  fitting a classification head while freezing the model parameters, including:
\textbf{GPT-GNN} \cite{hu2020gpt}, a GNN pre-training approach by a self-supervised graph generation task;
\textbf{DGI} \cite{velickovic2019deep}, a GNN pre-training approach that maximizes the mutual information between local- and global-level representations; \textbf{SAGE$_\text{self}$} \cite{hamilton2017inductive}, the self-supervised version of GraphSAGE, encouraging similar embeddings for neighboring nodes and distinct embeddings for non-adjacent nodes. Both DGI and SAGE$_\text{self}$ freeze the pre-trained weights, and fit a logistic regression model for the downstream classification.

(3) \emph{Pre-trained transformers}, which are pre-trained using masked language modeling, and then are fine-tuned together with a randomly initialized classification head (\eg, a fully connected layer) for the downstream classification, including:
\textbf{BERT} \cite{kenton2019bert}; \textbf{RoBERTa} \cite{liu2019roberta}; \textbf{BERT$^{*}$} and \textbf{RoBERTa$^{*}$}, variants of BERT and RoBERTa, which are obtained by further pre-training the pre-trained BERT and RoBERTa, respectively, using masked language modeling  on our datasets, to mitigate the domain gap between our datasets and the corpus used for pre-training BERT and RoBERTa.

(4) \emph{Prompt tuning}: \textbf{P-Tuning v2} \cite{liu-etal-2022-p}, is a version of prefix-tuning \cite{li2021prefix} optimized and adapted for natural language. It uses deep prompt tuning, which applies continuous prompts for every layer of the pre-trained language model. 

Note that our setting is distinct from few-shot learning under the meta-learning paradigm \cite{finn2017model}, as there is no  few-shot tasks for the meta-training phase. Hence, we cannot use state-of-the-art meta-learning models for comparison. Besides, two of the baselines we compared,  DGI and SAGE$_\text{self}$, have adopted a form of linear probe which is known to be a strong few-shot learner \cite{tian2020rethinking}.

\stitle{Baselines for zero-shot classification.}
We only compare with decoder-based LLMs and encoder-based PLMs, as all the other methods require at least one shot to work. 
Firstly, for LLMs with fewer than 10 billion parameters, we used Qwen-7B \cite{bai2023qwen}, BLOOMz-7B \cite{muennighoff2022crosslingual}, and the most capable openly available LLM to date, Llama-3-8B \cite{llama3}. For models with more than 10 billion parameters, we utilized Baichuan-13B \cite{baichuan13b}, which is currently the open-source model with the most training data at the 13 billion parameter scale. And we used Vicuna-13B \cite{zheng2024judging}, which matches over 90\% of the quality of OpenAI's ChatGPT \cite{achiam2023gpt} and Google Bard \cite{bard} while outperforming models like LLaMA and Stanford Alpaca \cite{Alpaca} in over 90\% of cases.
Secondly, for each encoder-based PLM, we use the discrete prompt $\mathtt{[CLASS]}$ (\ie, the label text alone).
We also evaluate handcrafted prompts  ``$\mathtt{prompt\ [CLASS]}$'', 
where $\mathtt{prompt}$ is a sequence of tokens found by prompt engineering, and annotate the model name with ``+d''.
We compute the   similarity between the target document and the label text of each class (with or without $\mathtt{prompt}$), and predict the most similar class following Fig.~\ref{fig:zero-shot}.

\stitle{Parameter settings.}
For \model, the text encoder is a transformer \cite{vaswani2017attention}. Following CLIP \cite{radford2021learning}, we use a 63M-parameter, 12-layer 512-wide model with 8 attention heads. It operates on a lower-cased byte pair encoding (BPE) representation of the texts with a 49,152 vocabulary size \cite{sennrich2016neural}. The max sequence length is capped at 128. The graph encoder employs a GCN \cite{KipfW17}, using two layers \cite{hamilton2017inductive} with a LeakyReLU activation, each with 128 dimensions \cite{perozzi2014deepwalk}.
The pre-training of our model starts from scratch without initializing the graph and text encoders with previously pre-trained weights.
$\lambda$ in Eq.~\eqref{eq:final loss} is set to $0.1$ on Cora, and set to $10$ on the three amazon review datasets, which were chosen from \{0.01, 0.1, 1, 10, 100\} according to the accuracy on validation data. The number of learnable prompt tokens, $M$ in Eq.~\eqref{eq:prompt embeddings}, is set to $4$, which was chosen from \{2, 4, 8, 16, 32\} according to the  accuracy on validation data. We use the Adam optimizer with the learning rate $2\times 10^{-5}$ with 2 training epochs, and a batch size of 64 in pre-training, referring to Hugging Face's \cite{wolf2020transformers} example settings. The text embedding size is 128, same to the output from the graph encoder.
To generate the summary embedding and the context-based prompt initialization, the number of neighboring nodes sampled
is 3.
For prompt tuning, we set the learning rate as $0.01$, which was chosen from \{0.0001,0.001,0.01,0.1\} according to the accuracy on validation data.  
Lastly, the Meta-net of \model$^*$ is built with a two-layer bottleneck structure (Linear-ReLU-Linear), with the hidden-layer dimension being 8.



\subsection{Performance of \model}
\label{sec:expt:g2p2}

We evaluate the performance under various few-shot settings. 

\stitle{Five-shot setting.} In Tab.~\ref{table:few-shot}, we first compare the performance of \model\ with baselines under the \emph{5-shot} setting. \model\ emerges as the winner consistently, outperforming the best baseline by around 2--8\% with statistical significance.

We also make a few more observations.
Firstly, among the GNNs, pre-trained/self-supervised models tend to perform better than the end-to-end approaches, since the latter heavily rely on labeled data. 
Among the former, DGI and SAGE$_\text{self}$ perform better as they are a form of linear probe, known to be a strong few-shot learner \cite{tian2020rethinking}.
Note that, instead of using word2vec embeddings \cite{mikolov2013efficient} of raw texts as node features, we also tired using the pre-trained RoBERTa \cite{liu2019roberta} to generate the node features for  DGI and SAGE$_\text{self}$.
However, doing so does not bring any improvement, showing that it is ineffective to simply combine a language model and GNN in a decoupled manner. In contrast, our proposed model jointly learns the text and graph encoders through three graph-grounded contrastive strategies. 
%
Secondly, PLMs are generally superior to GNNs, illustrating the importance of leveraging texts in a fine-grained way. Additionally, RoBERTa outperforms BERT owing to an improved pre-training procedure \cite{liu2019roberta}. However, further training PLMs on our texts gives mixed results: 
RoBERTa$^*$ slightly outperforms RoBERTa but BERT$^*$ is much worse than BERT. That means it is not straightforward to mitigate the domain gap by simply continuing training on the domain texts.
%
Thirdly,  the continuous prompt approach P-Tuning v2 achieves competitive results compared to fine-tuning, while having the advantage of being much cheaper than fine-tuning. 
Nevertheless, it is still significantly outperformed by our model \model. 
Furthermore, \model-p without prompt tuning is inferior to \model, showing the benefit of continuous prompts.

\begin{figure}
   \subfigure[\# of shots, Cora]{
   \centering
   \includegraphics[width=0.48\linewidth]{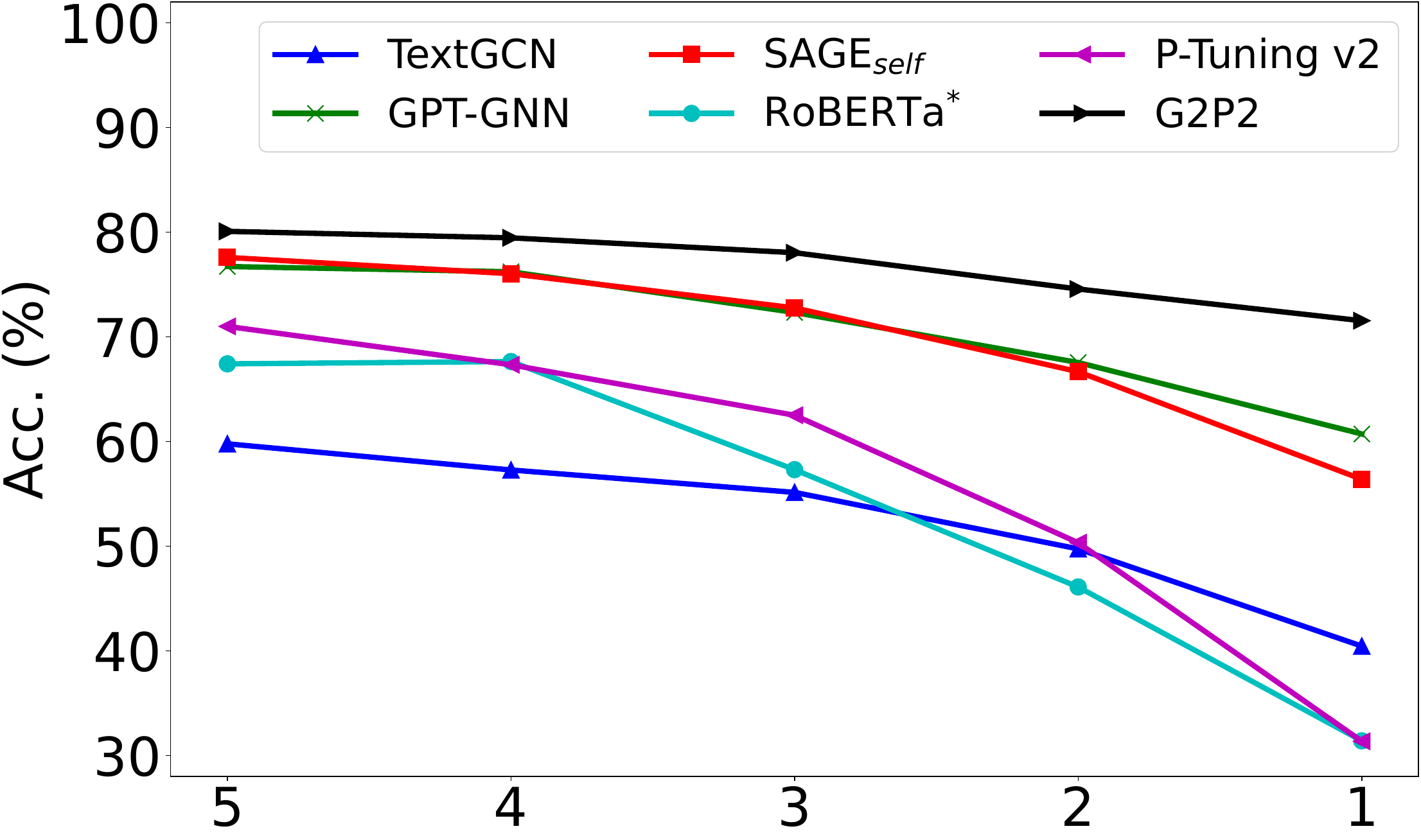}
   }%
   \subfigure[\# of shots, Art]{
   \centering
   \includegraphics[width=0.48\linewidth]{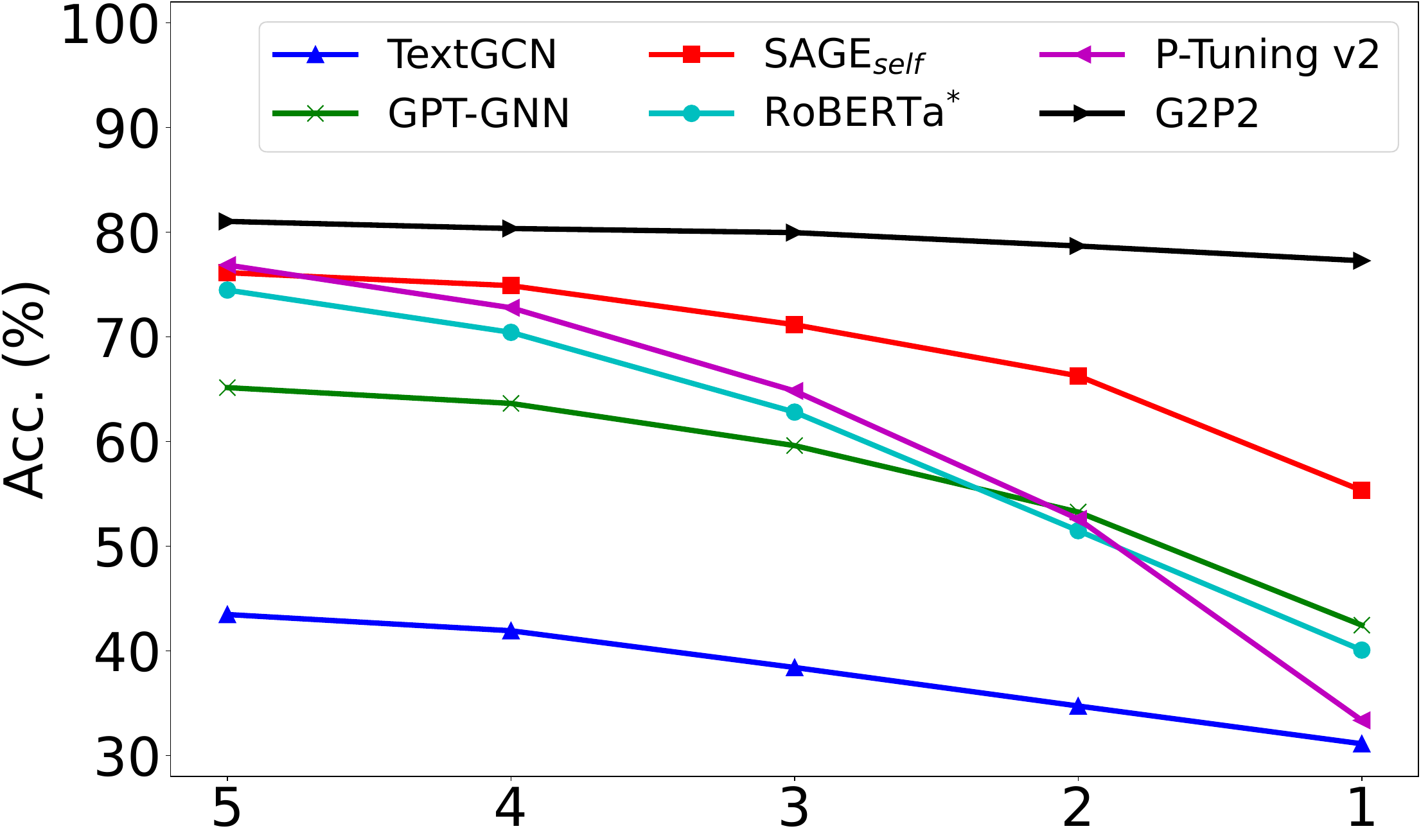}}
   \subfigure[\# of shots, Industrial]{
   \centering
   \includegraphics[width=0.48\linewidth]{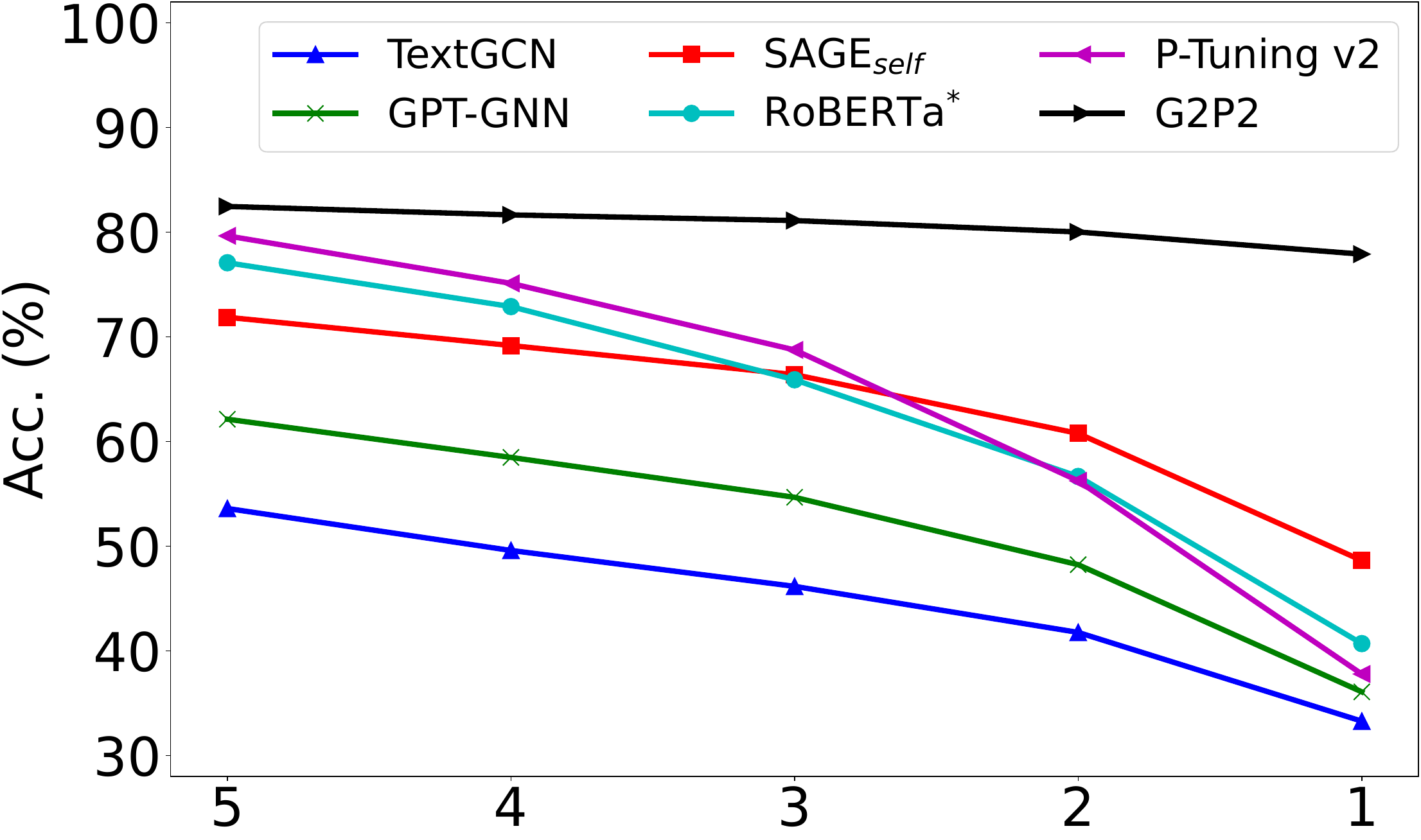}}%
   \subfigure[\# of shots, M.I.]{
   \centering
   \hspace{1mm}%
   \includegraphics[width=0.48\linewidth]{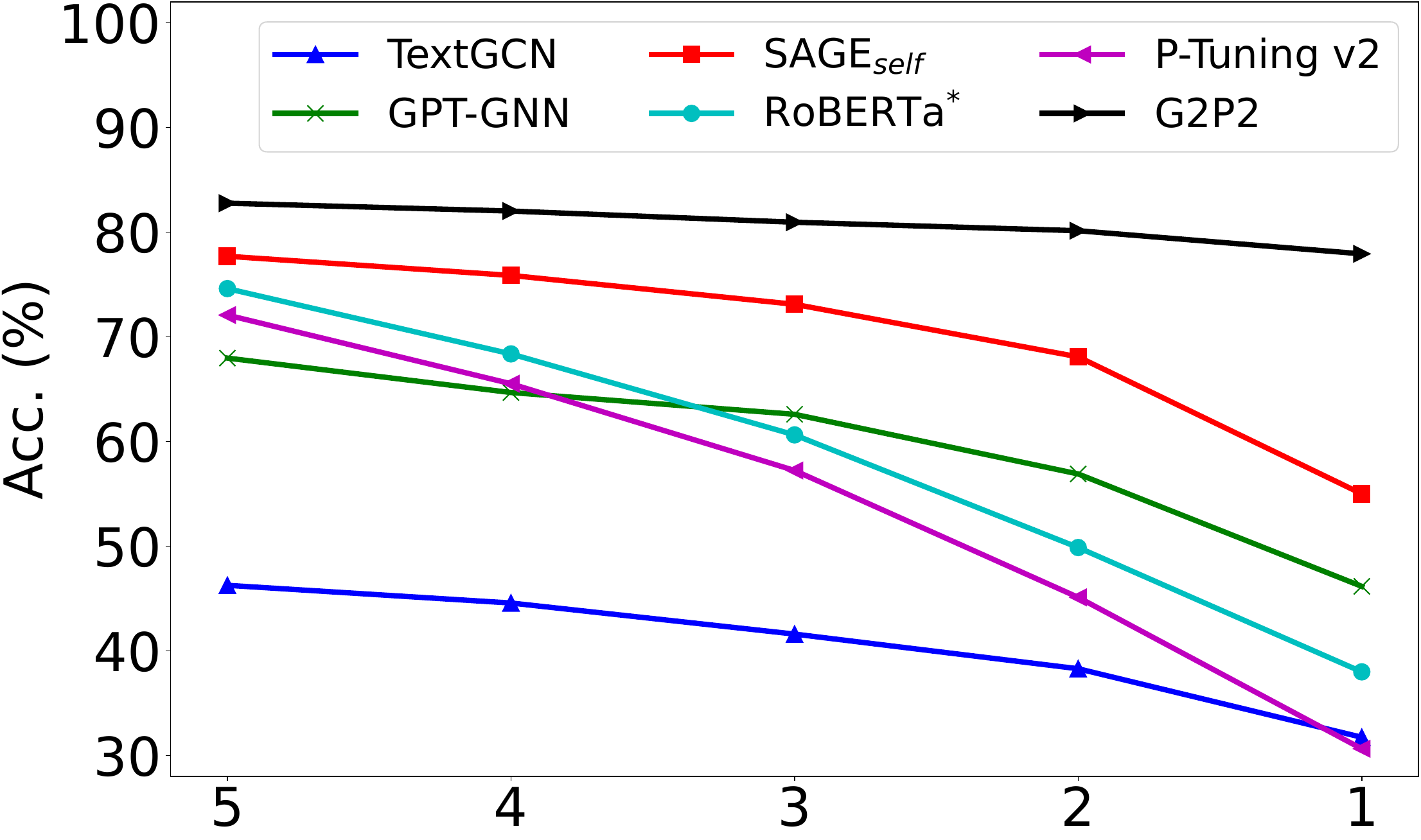}
   }
   \vspace{-3mm}
	\caption{Classification performance on different shots.}
	\vspace{-4mm}
	\label{fig:shots}
\end{figure}

\stitle{Settings with fewer shots.}
In addition to the 5-shot setting, in Fig.~\ref{fig:shots} we also study the impact of fewer shots  on \model\ and several representative baselines. \model\ generally performs the best across different shots. 
In general, the performances of all approaches degrade as fewer shots become available. 
However, the baselines suffer significantly under extreme low-resource (\eg, 1- or 2-shot) settings. In contrast, \model\ remains robust, reporting a relatively small decrease in performance even with just 1 or 2 shots.

It demonstrates the practical value of our proposed model especially when labeled data are difficult or costly to obtain in time. On the other hand, 
traditional approaches constantly face the challenge of the inability to keep up with the rapid growth of emerging classes in dynamic and open environments \cite{wang2021zero}.  For example, labeling a large volume of texts for novel topics in online articles, or new product categories in open-ended e-commerce platforms, can suffer a substantial time lag.


\stitle{Zero-shot setting inference.}
Finally, we report the zero-shot performance in Tab.~\ref{table:zero-shot}, where our models \model\ and \model+d significantly outperforms both the decoder-based LLMs and encoder-based PLMs baselines. The results particularly demonstrate the effectiveness of our graph-grounded contrastive pre-training in the absence of labeled data, which is crucial to handling evolving classes without any labeled sample in many real-world scenarios. 
We make several further observations.

Firstly, we specifically utilized five state-of-the-art decoder-based LLMs. 
Empirically, those LLMs requires lots of computational resources and time, \eg, a 13B model requires \textbf{more than 32GB} of GPU memory for inference with a \textbf{batch size of 1}. And it takes \textbf{over two weeks} for a single LLM to complete \textbf{one inference run} on the Art test set. In contrast, an encoder-based model like our G2P2 takes less than five hours to complete five inference runs on the largest Art test set.
As shown in Tab.~\ref{table:zero-shot}: (1) while decoder-based LLMs provide useful insights, their computational requirements are significantly higher, making them less efficient for the given task; (2) although decoder-based LLMs are famous for zero-shot scenarios, our G2P2 model outperforms them in these settings, providing more accurate results with lower computational overhead.

Secondly, for results about decoder-based PLMs, we can see that handcrafted discrete prompts (\ie, BERT$^*$+d and \model+d) can be superior to using label text only (\ie, BERT$^*$ and \model), showing the effectiveness of additional prompt tokens. 

However, finding the optimal discrete prompts often requires significant engineering work. Specifically, for the three approaches with discrete prompts, namely, RoBERTa$^{*}$+d, BERT$^{*}$+d and \model+d, we have explored more than 10 handcrafted prompt templates on each dataset,
which are typically specific to the dataset and require some domain knowledge to devise.
While discrete prompts are generally helpful, their effectiveness varies. 
For example, on the Cora dataset, while ``$\mathtt{a\ model\ of\ [CLASS]}$" is the best prompt for RoBERTa$^{*}$+d, it is a bad choice for \model+d. Sometimes, generic prompts like ``$\mathtt{a\ [CLASS]}$" can be the best choice.
Note that in Tab.~\ref{table:zero-shot}, we simply report the performance of the best template for each approach and dataset.
Hence, using the label text only is still a reasonably good choice.

\stitle{Zero-shot case study.}
To better understand the motivation and strength of G2P2, we conduct an zero-shot case study, as illustrated in Tab.~\ref{tab:quality}. We randomly selected a testing sample from a 5-way task in the Cora test set, and both the original document and label text are presented in the table.

Decoder-based LLMs cannot provide accurate classification results when only given the original document and label text. Therefore, after experimenting with various prompts, we crafted the instructions shown in blue in the table.
It's evident that zero-shot classification tasks remain challenging for popular decoder-based LLMs. In this example, four of the five label texts are semantically similar, making them difficult to differentiate. Only BLOOMz provided the correct answer, while the other four models produced various errors. Qwen generated irrelevant content, Llama-3 repeated the question without providing reasons, and Vicuna entered a repetitive loop—a common issue with decoder-based LLMs.

In contrast, our encoder-based models are more accurate and reliable due to directly using embeddings containing fine-grained information and computing cosine similarity. They are also more time- and computationally efficient due to having fewer parameters than decoder-based LLMs. Among these encoder-based models, G2P2 not only selected the correct answer but also showed a clear distinction between the similarities of the correct and incorrect answers.
This is because using GNNs to encode graph structures expands the usable information from just text to both text and graph structures. Our model effectively leverages these dual modalities to tackle the zero-shot text classification challenge.

\begin{table}[tbp]
    \footnotesize
	\centering 
 	\addtolength{\tabcolsep}{-5pt}
	\caption{\emph{Zero-shot} classification accuracy (percent).  
	} 
	\label{table:zero-shot} 
   {
    \vspace{-2mm}
   \scriptsize
   }
	\begin{tabular}{@{}c|c|c|c|c|c|c@{}}  
		\toprule
		  &Models &\# Params. &Cora&Art&Industrial&M.I.
          \\\midrule
          \multirow{5}{*}{Decoder} 
          &Qwen&7B&21.67&18.65&19.13&19.00 \\
          &BLOOMz&7B&\underline{48.89}&23.60&24.07&20.06 \\
          &Llama-3&8B&25.32&18.94&19.70&19.47 \\
          &Baichuan&13B&21.67&16.68&19.22&18.68 \\
          &Vicuna&13B&16.65&15.36&16.81&15.85 

		 \\\midrule
          \multirow{9}{*}{Encoder} 
		 & RoBERTa &123M &30.46 &42.80 &42.89 &36.40 \\
         & RoBERTa$^{*}$ &123M &39.58 &34.77 &37.78 &32.17 \\
        & RoBERTa$^{*}$+d &123M &45.53 &36.11 &39.40 &37.65 \\
		& BERT &110M &23.58 &35.88 &37.32 &37.42 \\
        & BERT$^{*}$ &110M &23.38 &54.27 &\underline{56.02} &50.19 \\
        & BERT$^{*}$+d &110M &26.65 &\underline{56.61} &55.93 &\underline{52.13} \\
		& \model &66M &64.75 &76.62 &76.43 &74.44 \\
		& \model+d &66M & \textbf{66.43}$^{*}${}   & \textbf{76.95}$^{*}${} 
   & \textbf{77.31}$^{*}${} & \textbf{75.94}$^{*}${}\\
		& (improv.) &\--  & (+35.88\%)& (+35.93\%)&(+38.00\%)& (+45.67\%)\\
	\bottomrule
	\end{tabular}
	\vspace{-4mm}
\end{table}

\begin{table*}[t]
\small
\centering 
\caption{\emph{Zero-shot} case study. An example from Cora, with the ground-truth answer \underline{\texttt{\textbf{B}}}. For encoder-based PLMs, the cosine similarity here is softmax normalised. 
\emph{Italic} text is the instruction crafted by us for decoder-based LLMs.}
\vspace{-2mm}
\renewcommand{\arraystretch}{0.4}
\begin{tabular}{@{}p{2.9cm}|p{15cm}@{}}
		\toprule
		 \multirow{1}{*}{\textbf{Document}}  & \parbox[c]{15cm}{\setstretch{0.4}\texttt{this paper develops a new and natural parallel vector model, and shows that for all k 1, the languages recognizable in o(log k n) time and polynomial work in the model are exactly those in nc k . some improvements to other simulations in the literature of parallel models and reversal complexity are given.}} \\\midrule
       \multirow{8}{*}{\textbf{Label text}}  
       &\texttt{(A) data structures  algorithms and theory, computational geometry}  \\
       &\texttt{(B) data structures  algorithms and theory, computational complexity }  \\
       &\texttt{(C) data structures  algorithms and theory, parallel }  \\
       &\texttt{(D) data structures  algorithms and theory, logic }  \\
       &\texttt{(E) artificial intelligence, nlp}  \\\midrule
        \multicolumn{2}{c}{Decoder-based LLMs} \\\midrule

    \multirow{14}{*}{\textbf{Prompt for}} 
    & \texttt{\emph{This is a paper abstraction:}} \\               
    \multirow{12}{*}{\textbf{decoder-based}} 
    & \parbox[c]{15cm}{\setstretch{0.4}\texttt{this paper develops a new and natural parallel vector model, and shows that for all k 1, the languages recognizable in o(log k n) time and polynomial work in the model are exactly those in nc k . some improvements to other simulations in the literature of parallel models and reversal complexity are given.}} \\ 
    \multirow{1}{*}{\textbf{LLMs}}
    & \parbox[c]{15cm}{\setstretch{0.4}\texttt{\emph{Classify it into one of the five categories: (A) data structures  algorithms and theory, computational geometry  (B) data structures  algorithms and theory, computational complexity (C) data structures  algorithms and theory, parallel (D) data structures  algorithms and theory, logic (E) artificial intelligence, nlp. The answer can only be (A), (B), (C), (D), or (E).}}} \\
    & \texttt{\emph{\#\#\# Answer Choice:}} \\
    \midrule

       \textbf{Qwen (\text{7B})}   & \texttt{\textbf{A}, B, C, D, or E. for example in examples: if example in lines: ans = example} \\
            \midrule
		\textbf{BLOOMz (\text{7B})}  & \texttt{(\textbf{B}) data structures  algorithms and theory, computational complexity</s>} \\\midrule
  
            \multirow{12}{*}{\textbf{Llama-3 (\text{8B})}} 
            &\texttt{\textbf{A} \#\#\# Explanation:} \\
            &\texttt{This is a paper abstraction:} \\
            &
            \parbox[c]{15cm}{\setstretch{0.4}\texttt{this paper develops a new and natural parallel vector model, and shows that for all k 1, the languages recognizable in o(log k n) time and polynomial work in the model are exactly those in nc k . some improvements to other simulations in the literature of parallel models and reversal complexity are given.}}
            \\\midrule

            \textbf{Baichuan (\text{13B})}  &\texttt{The correct choice would be category \textbf{A}): Data Structures \& Algorithms Theory} \\
            \midrule

            
            \multirow{8}{*}{\textbf{Vicuna (\text{13B})}}  & \texttt{\# (\textbf{A}) data structures  algorithms and theory} \\
            & \texttt{\# (B) data structures  algorithms and theory} \\
            & \texttt{\# (C) data structures  algorithms and theory} \\
            & \texttt{\# (D) data structures  algorithms and theory} \\
            & \texttt{\# (E) data structures  algorithms and theory} \\
            \midrule

            \multicolumn{2}{c}{Encoder-based PLMs} \\\midrule
            


            \multirow{8}{*}{\textbf{RoBERTa$^{*}$+d (\text{123M})}}  
            & \texttt{ $<$Document embedding, embedding of Label A with discrete prompt$>$ = 0.1953 } \\
            & \texttt{ $<$Document embedding, embedding of Label \textbf{B} with discrete prompt$>$ = \textbf{0.2074} } \\
            & \texttt{ $<$Document embedding, embedding of Label C with discrete prompt$>$ = 0.2068 } \\
            & \texttt{ $<$Document embedding, embedding of Label D with discrete prompt$>$ = 0.1986 } \\
            & \texttt{ $<$Document embedding, embedding of Label E with discrete prompt$>$ = 0.1919} \\
            \midrule

            \multirow{8}{*}{\textbf{BERT$^{*}$+d (\text{110M})}}  
            & \texttt{ $<$Document embedding, embedding of Label A with discrete prompt$>$ = 0.1936} \\
            & \texttt{ $<$Document embedding, embedding of Label B with discrete prompt$>$ = 0.2040} \\
            & \texttt{ $<$Document embedding, embedding of Label \textbf{C} with discrete prompt$>$ = \textbf{0.2068}} \\
            & \texttt{ $<$Document embedding, embedding of Label D with discrete prompt$>$ = 0.2036} \\
            & \texttt{ $<$Document embedding, embedding of Label E with discrete prompt$>$ = 0.1919} \\
            \midrule

            \multirow{8}{*}{\textbf{G2P2+d (\text{63M})}}  
            & \texttt{ $<$Node embedding, embedding of Label A with discrete prompt$>$ = 0.2021} \\
            & \texttt{ $<$Node embedding, embedding of Label \textbf{B} with discrete prompt$>$ = \textbf{0.2245}} \\
            & \texttt{ $<$Node embedding, embedding of Label C with discrete prompt$>$ = 0.2100} \\
            & \texttt{ $<$Node embedding, embedding of Label D with discrete prompt$>$ = 0.2002} \\
            & \texttt{ $<$Node embedding, embedding of Label E with discrete prompt$>$ = 0.1633} \\
            
		\hline
\end{tabular}
\vspace{-4mm}
\label{tab:quality}
\end{table*}

\subsection{Model analyses of \model}

We conduct more in-depth studies on \model. By default, we report the classification \emph{accuracy} under the \emph{5-shot} setting.

\stitle{Ablation study.}
We first evaluate the contribution from each of the three interaction-based contrastive strategies, by employing different combinations of the proposed loss terms $\bL_1,\bL_2$ and $\bL_3$. 
As shown in Tab.~\ref{table:ablation}, strategies without $\bL_1$ have performed quite poorly, demonstrating that the bijective text-node interaction is the fundamental component of our pre-training. 
That being said, when further adding $\bL_2$ or $\bL_3$ to $\bL_1$, we still observe a noticeable performance improvement, showing the benefit of incorporating additional graph-based interactions for text data.
Lastly, \model\ with all three loss terms outperforms all 1- or 2-combinations of the losses, demonstrating that the three contrastive strategies are all useful and they are well integrated.
Overall, the results reveal that graph information is vital to low-resource text classification, since graph structures reveal rich relationships between documents.


Next, we evaluate the contribution from our prompt-tuning. Specifically, we compare \model\ with two ablated variants: using label text only without trainable prompt vectors, and randomly initializing the prompt vectors. As reported in Tab.~\ref{table:ablation}, only using label text clearly degrades the classification performance, implying the importance of learning continuous prompts via prompt tuning. Furthermore, our approach \model\ with context-based initialization for prompt vectors shows a small but consistent advantage over random initialization, implying the usefulness of considering graph structures in prompt tuning.

\begin{table}[tbp]
    \footnotesize 
	\centering 
 	\addtolength{\tabcolsep}{-0.1mm}
	\caption{Ablation study with variants of \model.  
	} 
	\label{table:ablation} 
 \vspace{-2mm}
	\begin{tabular}{@{}c|c|c|c|c@{}}  
		\toprule
		  &Cora&Art&Industrial&M.I.
		 \\\midrule
		 Only $\bL_3$  &74.66$\pm$1.80 &52.56$\pm$1.09 &45.97$\pm$0.81 &49.05$\pm$0.54\\
          Only $\bL_2$ &77.01$\pm$1.30 &58.90$\pm$0.55 &52.99$\pm$0.46 &59.41$\pm$0.85\\
          Only $\bL_1$ &79.50$\pm$1.19 &77.37$\pm$0.72 &78.10$\pm$0.34 &79.70$\pm$0.56\\
          $\bL_2$+$\bL_3$ &70.04$\pm$2.89 &49.91$\pm$1.57 &50.07$\pm$0.50 &56.14$\pm$1.01\\
          $\bL_1$+$\bL_3$ &79.73$\pm$0.89 &78.60$\pm$0.40 &79.97$\pm$0.43 &80.42$\pm$0.45\\
          $\bL_1$+$\bL_2$ &79.42$\pm$1.04 &80.55$\pm$0.52 &81.06$\pm$0.33 &82.39$\pm$0.41\\
          \midrule
          Only label text &79.16$\pm$1.23 &79.59$\pm$0.31 &80.86$\pm$0.40 &81.26$\pm$0.36\\
          Random init. &80.03$\pm$0.99 &80.85$\pm$0.43 &82.43$\pm$0.33 &82.64$\pm$0.21\\
		\midrule
		\model & \textbf{80.08}$\pm$1.33   & \textbf{81.03}$\pm$0.43  & \textbf{82.46}$\pm$0.35 & \textbf{82.77}$\pm$0.32 \\
	\bottomrule
	\end{tabular}
	\vspace{-2mm}
\end{table}

\begin{figure}
   \subfigure[Interaction coefficient, $\lambda$]{
   \centering
   \includegraphics[width=0.48\linewidth]{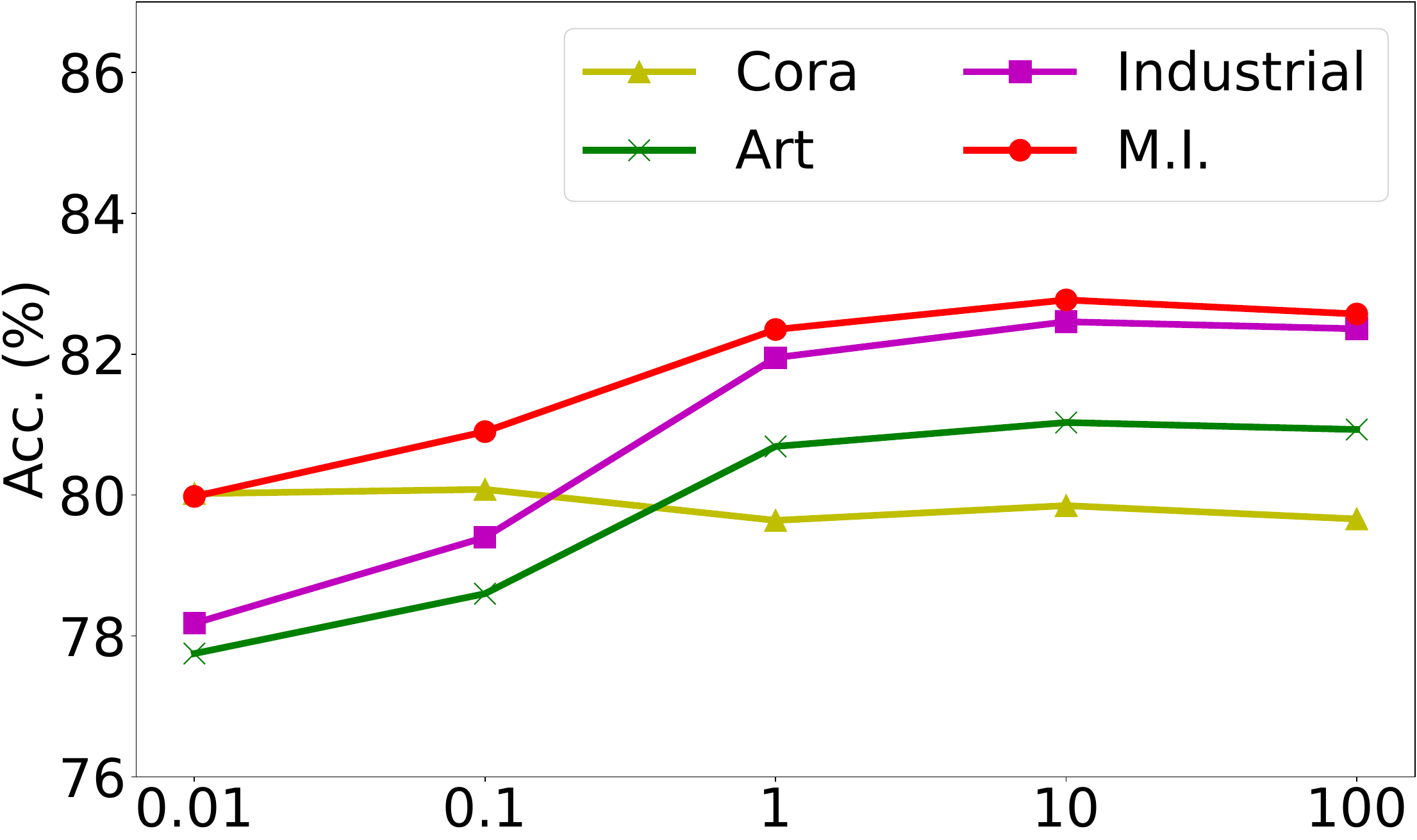}
   }
   \subfigure[Prompt length, $M$]{
   \centering
   \includegraphics[width=0.48\linewidth]{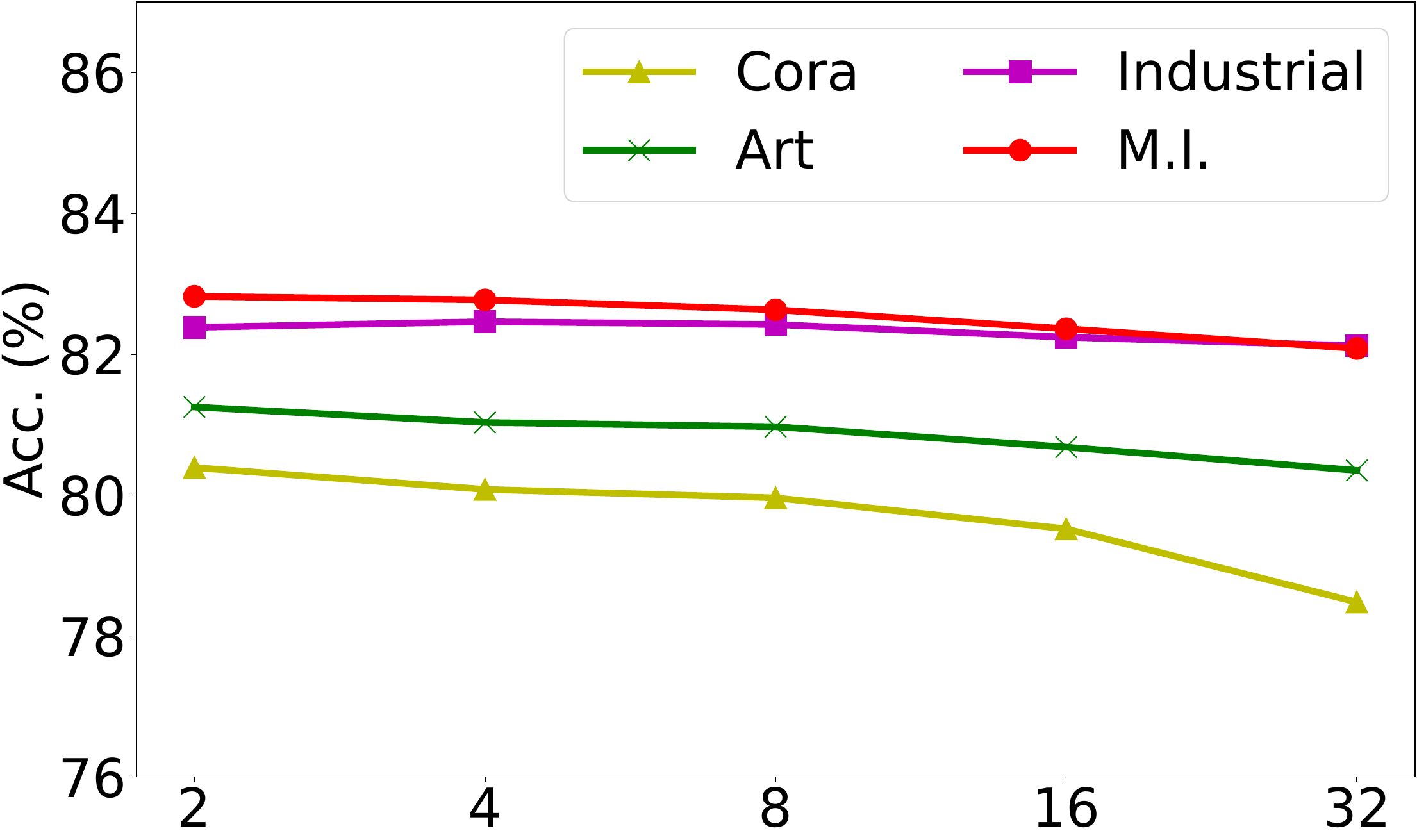}
   }
   \vspace{-2mm}
	\caption{Hyperparameter study for \model.}
	\label{fig:param}
	\vspace{-2mm} 
\end{figure}

\begin{table}[tbp]
	\centering 
     \footnotesize
 	\addtolength{\tabcolsep}{-0.9mm}
	\caption{Analysis of tuning time and parameter size. } 
	\label{table:efficiency} 
 \vspace{-2mm}
	\begin{tabular}{@{}c|c|c|c|c|c@{}}  
		\toprule
       & \multicolumn{4}{c|}{Tuning time per task (in seconds)} & Param. \\\cmidrule{2-5}
		  &Cora&Art&Industrial&M.I.&size
		 \\\midrule
		 RoBERTa & 45.47$\pm$2.38 & 64.22$\pm$3.62 & 43.46$\pm$2.99 & 44.99$\pm$2.58 & 123M\\
          RoBERTa$^{*}$ & 39.38$\pm$2.01 & 59.56$\pm$3.55 & 35.10$\pm$2.75 & 38.84$\pm$2.39 & 123M\\
		 BERT & 32.23$\pm$1.71 & 51.77$\pm$2.00 & 31.72$\pm$1.77 & 33.55$\pm$2.39 & 110M\\
          BERT$^{*}$ & 34.82$\pm$1.68 & 55.16$\pm$2.32 & 31.11$\pm$1.74 & 29.00$\pm$2.23 & 110M\\
		\midrule
		\model & \textbf{2.42}$\pm$0.41   & \textbf{22.03}$\pm$1.39 & \textbf{14.63}$\pm$1.26 & \textbf{12.72}$\pm$1.17 & \textbf{2048}\\
	\bottomrule
	\end{tabular}
\end{table}

\begin{table}[tbp]
	\centering 
     \footnotesize
 	\addtolength{\tabcolsep}{2pt}
	\caption{Inductive performance on text classification.} 
	\label{table:generalization} 
 \vspace{-2mm}
	\begin{tabular}{c|c|c|c}  
		\toprule
		  &Art&Industrial&M.I.
		 \\\midrule
          BERT$^{*}$ &43.66$\pm$0.90  &48.35$\pm$0.25 &39.24$\pm$0.88 \\
		 RoBERTa$^{*}$ &69.55$\pm$1.14  &73.65$\pm$0.86 &71.96$\pm$1.44 \\ 
		\midrule
		\model& \textbf{79.81}$\pm$0.22   & \textbf{81.29}$\pm$0.32
  & \textbf{81.85}$\pm$0.33  \\
	\bottomrule
	\end{tabular}
\vspace{-4mm}
\end{table}

\stitle{Hyperparameter study.}
We first investigate the impact of the interaction coefficient $\lambda$ in Fig.~\ref{fig:param}(a), which balances the high-order contrastive losses ($\bL_2,\bL_3$). The performance is generally better and stable when $\lambda$ is slightly bigger (\eg, $\ge 10$), indicating the significance of the high-order text-summary and node-summary interactions. 
Next, we study the prompt length $M$ in Fig.~\ref{fig:param}(b), which refers to the number of trainable prompt vectors in Sect.~\ref{sec:model_prompt}.
The performance is relatively unaffected by the prompt length, and thus it is robust to choose a small $M$ (\eg, 4) for efficiency.

\stitle{Efficiency of prompt tuning.}
In this experiment, we investigate the prompt tuning efficiency of \model\ in comparison to the efficiency of traditional fine-tuning. 
As \model\ has a transformer component, we compare it with four transformer based models, all of which follow the classical ``pre-train, fine-tune'' paradigm.

As shown in Tab.~\ref{table:efficiency},
prompt tuning in \model\ is much more efficient than fine-tuning in the baselines, achieving 2.1$\sim$18.8x speedups. The reason is that prompt tuning updates far fewer parameters. In \model, we used 4 trainable 512-dimensional prompt vectors, totalling to 2048 parameters only, while fine-tuning in the baselines needs to update the whole pre-trained model with more than 100M parameters. Note that the speedup is not linear w.r.t.~the parameter size, due to overheads in the data loader and the optimizer. 
Overall, our prompt tuning is not only effective under low-resource settings, but also parameter- and computation-efficient.

\stitle{Inductive ability.}
Our previous experiments can be deemed transductive as both the pre-training and downstream text classification are conducted on the whole corpus.
To further evaluate the generalization ability of  \model, we adopt an ``inductive'' setting, whereby we pre-train the text encoder only on a subset of the corpus and perform downstream classification on a disjoint subset. Particularly, in the three Amazon datasets, since user texts have no labels and item texts have labels, it is natural for us to pre-train with only user texts and classify only item texts downstream. We also employ masked language modeling on only the user texts for BERT and RoBERTa, to get BERT$^{*}$ and RoBERTa$^{*}$. As shown in Tab.~\ref{table:generalization}, \model\ still performs very well in the inductive setting, illustrating the strong generalization ability of our pre-trained model.

\begin{table*}[tbp]
    \footnotesize
    \centering
   	\caption{Performance on base and unseen classes within the same domain.}
        \vspace{-2mm}
    {
    \scriptsize Prompts are selected or tuned on the labeled set of the base classes, and tested on the test set of the base classes, or the unseen classes. HM denotes the\\ \emph{harmonic mean} of the testing performance on the base and unseen classes, enabling the assessment of the generalization trade-off between them \cite{xian2017zero}.
    }\\[0.5mm]
    \subfigure[Overall]{
   \centering
   \addtolength{\tabcolsep}{0pt}
   \begin{tabular}[width=1\linewidth]{l|c|c|c}
        \toprule
        & Base & Unseen & HM \\
    \midrule
    G2P2+d & 39.3 & \underline{38.0} & 38.4 \\
    G2P2  & \textbf{49.7} & 33.5 & \underline{39.5}\\
    \model$^*$ & \underline{48.7}& \textbf{41.0}& \textbf{43.9}\\
        \bottomrule
        \end{tabular}}
   \subfigure[Cora]{
   \centering
   \addtolength{\tabcolsep}{-4pt}
   \begin{tabular}[width=1\linewidth]{l|c|c|c}
        \toprule
        & Base & Unseen & HM \\
        \midrule
        G2P2+d & 28.81$\pm$3.35 & 22.51$\pm$1.41 & 25.14$\pm$1.42 \\
        G2P2  & \textbf{52.50}$\pm$1.38 & \underline{24.83}$\pm$2.71 & \textbf{33.65}$\pm$2.72 \\
        \model$^*$ & \underline{48.60}$\pm$1.68 &\textbf{25.70}$\pm$1.25 & \underline{33.62}$\pm$1.46 \\

        \bottomrule
        \end{tabular}
   }
   \subfigure[Art]{
   \centering
   \addtolength{\tabcolsep}{-4pt}
   \begin{tabular}[width=1\linewidth]{l|c|c|c}
        \toprule
        & Base & Unseen & HM \\
        \midrule
        G2P2+d &40.04$\pm$2.35 &\underline{38.01}$\pm$0.74  & \underline{38.95}$\pm$1.05 \\
        G2P2  &\underline{43.47}$\pm$2.59  &31.97$\pm$4.69  & 36.71$\pm$3.73 \\
        \model$^*$ &\textbf{44.29}$\pm$2.30  &\textbf{41.43}$\pm$1.28  & \textbf{42.80}$\pm$1.61 \\

        \bottomrule
        \end{tabular}
   }
   \subfigure[Industrial]{
   \centering
   \addtolength{\tabcolsep}{-2pt}
   \begin{tabular}[width=1\linewidth]{l|c|c|c}
        \toprule
        & Base & Unseen & HM \\
        \midrule
        G2P2+d &47.24$\pm$2.71 &\textbf{51.80}$\pm$1.66 & \underline{49.36}$\pm$1.64 \\
        G2P2  &\textbf{52.84}$\pm$1.72 &39.71$\pm$1.66 & 45.29$\pm$0.98 \\
        \model$^*$ &\underline{52.44}$\pm$2.48 &\underline{50.77}$\pm$1.74 & \textbf{51.59}$\pm$2.09 \\

        \bottomrule
        \end{tabular}
   }
   \subfigure[M.I.]{
   \centering
   \addtolength{\tabcolsep}{-2pt}
   \begin{tabular}[width=1\linewidth]{l|c|c|c}
        \toprule
        & Base & Unseen & HM \\
        \midrule
        G2P2+d &41.12$\pm$3.63  &\underline{39.64}$\pm$3.88  & 40.03$\pm$0.78 \\
        G2P2  &\textbf{49.89}$\pm$4.76 &37.49$\pm$3.95  & \underline{42.43}$\pm$1.64 \\
        \model$^*$ &\underline{49.29}$\pm$4.25  &\textbf{45.95}$\pm$1.92  & \textbf{47.47}$\pm$2.42 \\
   
        \bottomrule
        \end{tabular}

   }



	\label{fig: base and new}
	\vspace{-2mm} 
\end{table*}

\begin{table}[tbp]
    \footnotesize
	\centering 
        \vspace{-2mm}
	\caption{Continual learning within the same domain.}	\label{table:mixed classes} 
 \vspace{-2mm}
 {\scriptsize Prompts are selected or tuned on the labeled data from base classes, and\\ tested on the combined base classes (test set only) and unseen classes. 
 \\[1.5mm]
	\begin{tabular}{l|c|c|c|c}  
		\toprule
		  &Cora&Art&Industrial & M.I.
		 \\\midrule
          G2P2+d &18.02$\pm$1.70 &\underline{29.69}$\pm$1.27 &\underline{39.27}$\pm$0.28 &27.57$\pm$0.61  \\
		 G2P2  &\underline{23.53}$\pm$1.19 &27.45$\pm$2.33 &35.85$\pm$1.06 &\underline{32.17}$\pm$0.74 \\ 
	   \model$^*$ &\textbf{28.93}$\pm$1.42 &\textbf{31.09}$\pm$0.61 &\textbf{41.00}$\pm$1.69 &\textbf{34.66}$\pm$2.03 \\ 
   
	\bottomrule
	\end{tabular}
}
\vspace{-2mm}
\end{table}

\subsection{Performance and model analyses of \model$^*$}\label{sec:expt:conditional}

Finally, we conduct additional experiments to investigate the performance of \model$^*$, in particular the generalization ability of conditional prompt tuning to handle wider unseen classes. In these experiments, on each dataset we randomly sample some (35 for Cora, 66 for Art, 41 for Industrial, and 35 for M.I.) classes as the base classes, and the same number of classes as unseen classes. Each base class has five labeled instances (5-shot) for selecting or tuning the optimal prompts, and reserves the remaining instances for testing. In contrast, all instances of each unseen class are used for testing, without any labeled data for learning. We compare the three variants of our proposed model: (1) \model+d, the zero-shot method with handcrafted discrete prompts chosen based on the base classes; (2) \model, which learns continuous and static prompts from the base classes; (3) \model$^*$, which learns conditional prompts on the base classes.

In the following, we first conduct experiments to evaluate the  within- and cross-domain generalization capability. Then, we investigate the impact of hidden dimension of Meta-net, and the effect of increasing the parameters of \model.

\stitle{Within-domain generalization.}
Within the same domain, we consider two scenarios of generalization: from base classes to unseen classes, and a continual learning setting where base classes are still included during testing.

Firstly, we report the results on the testing sets of both the base classes and unseen classes in Tab.~\ref{fig: base and new}.
 Note that the base and unseen classes come from the same domain, \ie, the same dataset in our context. 
Specifically, on the base classes, \model+d performs the worst in all cases, showing that the discrete prompts are difficult to select for intrinsically continuous PLMs. In contrast, \model\ generally obtains the best performance on the base classes, showing that continuous prompt tuning can find better prompts if some labeled data are available for tuning. However, the learned static prompts do not generalize well to the unseen classes, showing a significant class shift between base and unseen classes. Interestingly, \model+d performs better on unseen classes than \model, showing that the handcrafted prompts are robust to class shifts to some extent. 
G2P2*, does not achieve the best performance on the base class of the \emph{Cora} dataset but is still the second best, outperforming the third-place G2P2+d by 68.7\%. Moreover, we achieve the best results on the new classes. As a result, the harmonic mean of our performance and that of the best-performing G2P2 shows no  difference.
On the \emph{Industrial} dataset, G2P2* is not the best performer on either the base class or the new class. However, it demonstrates balanced performance, ranking just below the best in both categories. Consequently, its overall performance, represented by the harmonic mean, is the highest, surpassing the second-best by 4.5\%.
Overall, in comparison to \model+d and \model, \model$^*$ demonstrates competitive performance on the base classes and excellent performance on the unseen  classes, implying that conditional prompt tuning not only fits well to the base classes, but also generalizes to the unseen classes.

Secondly, we address a more pragmatic ``continual learning'' scenario \cite{parisi2019continual}, in which we combine the test sets of both the base and unseen classes during inference. This scenario is designed to evaluate the undesirable effect of catastrophic forgetting of the base classes while adapting to the unseen classes. 
The results are presented in Tab.~\ref{table:mixed classes}. It becomes evident that \model\ loses its competitive edge over \model+d,
but \model$^*$ remains superior, implying that \model$^*$ is robust to continual learning with less ``forgetting'' of the base classes.

\begin{table*}[tbp]
\footnotesize
    \centering
    \addtolength{\tabcolsep}{1mm}
    \caption{Cross-domain generalization from Amazon to Cora.}\label{table:cross data} 
    \vspace{-2mm}
    {\scriptsize
    \emph{No generalization}: Prompts are tuned and tested on the labeled data and test set, respectively, of the base classes from the same domain.\\ \emph{Cross-domain generalization}: Prompts are tuned in the same way on the source domain, but tested on the unseen classes of the target domain.
    }
    \\[1.5mm]
    \begin{tabular}{l|ccc|ccc}
        \toprule
        & \multicolumn{3}{c|}{No generalization (base classes)} & \multicolumn{3}{c}{Cross-domain generalization (unseen classes)}\\\midrule
        Source (Tuning) & Art & Industrial & M.I. & Art & Industrial & M.I. \\\midrule
        Target (Testing) & Art & Industrial & M.I. & Cora & Cora & Cora  \\
        \midrule
        G2P2    & 43.47$\pm$2.59 & \textbf{52.84}$\pm$1.72 & \textbf{49.89}$\pm$4.76 & 22.83$\pm$4.63 & 20.88$\pm$3.99 & 19.51$\pm$5.08  \\
        \model$^*$       & \textbf{44.29}$\pm$2.30 & 52.44$\pm$2.48 & 49.29$\pm$4.25 & \textbf{27.68}$\pm$3.67 & \textbf{21.01}$\pm$3.94 & \textbf{27.05}$\pm$3.25 \\

        \bottomrule
    \end{tabular}
    \vspace{-2mm}
\end{table*}

\stitle{Cross-domain generalization.}
In the preceding part, we have investigated the ability to generalize to unseen classes within the same domain. In the following, we aim to examine the generalization to unseen classes across different domains. The ability to generalize to out-of-distribution data is a crucial feature in real-world scenarios, and it is intriguing to see whether the handcrafted or learned prompts can withstand shifts across domains. 
More specifically, we consider cross-domain scenarios within the Amazon ecosystem, and the scenarios with a more challenging shift from Amazon to Cora.

\begin{figure}[tbp]
   \subfigure[G2P2+d]{
   \centering
   \includegraphics[width=0.3\linewidth]{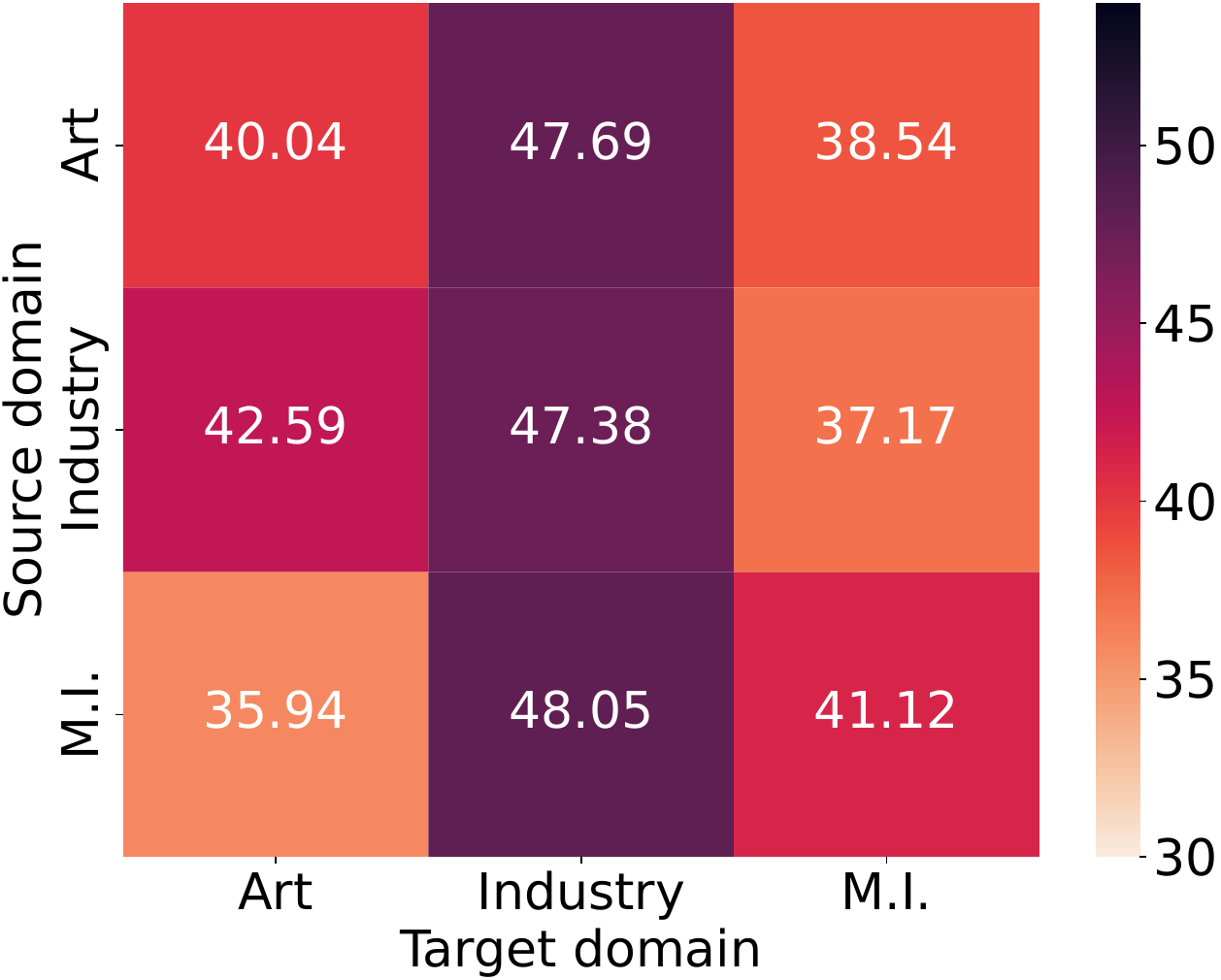}
   }
   \subfigure[G2P2]{
   \centering
   \includegraphics[width=0.3\linewidth]{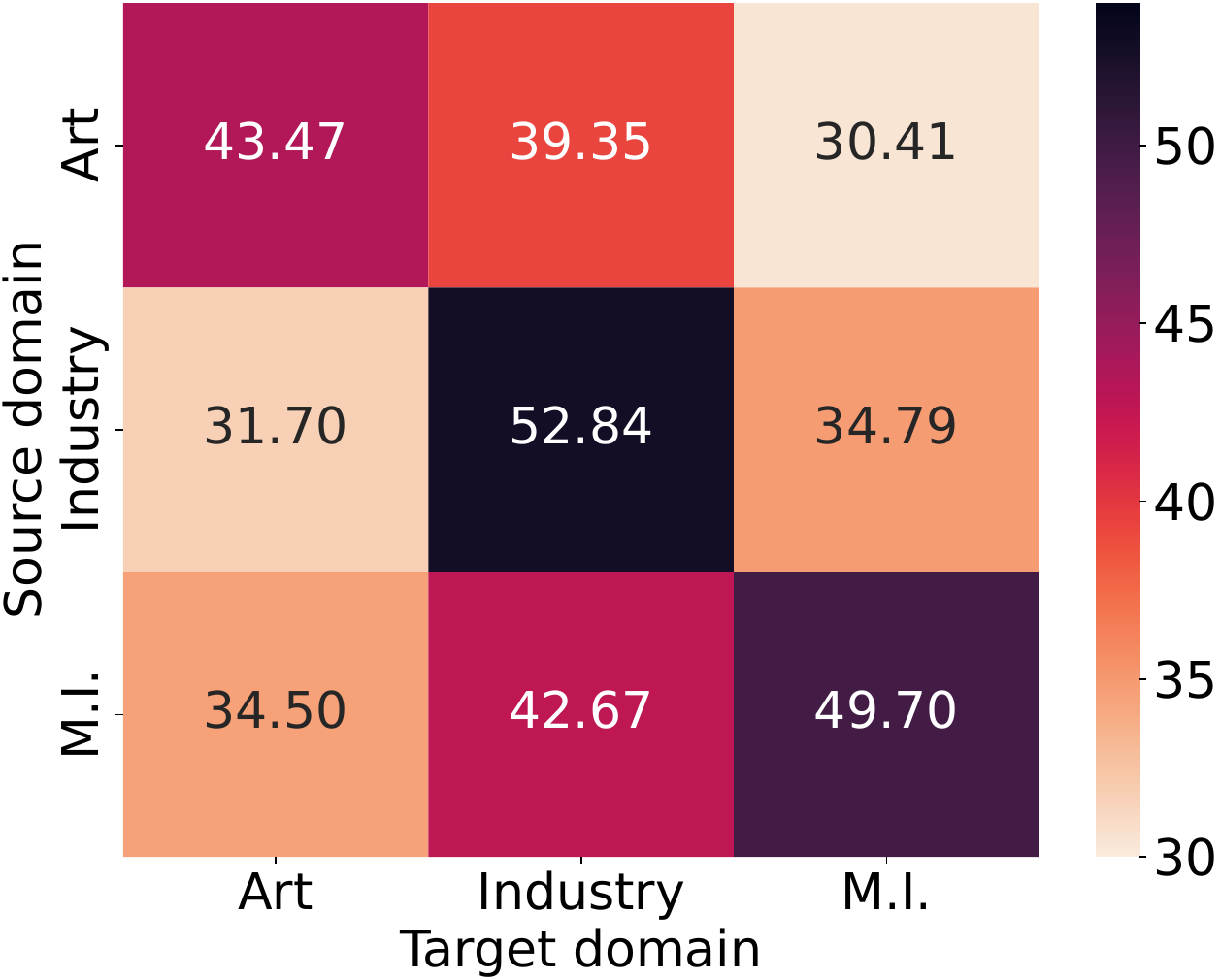}
   }
   \subfigure[\model$^*$]{
   \centering
   \includegraphics[width=0.3\linewidth]{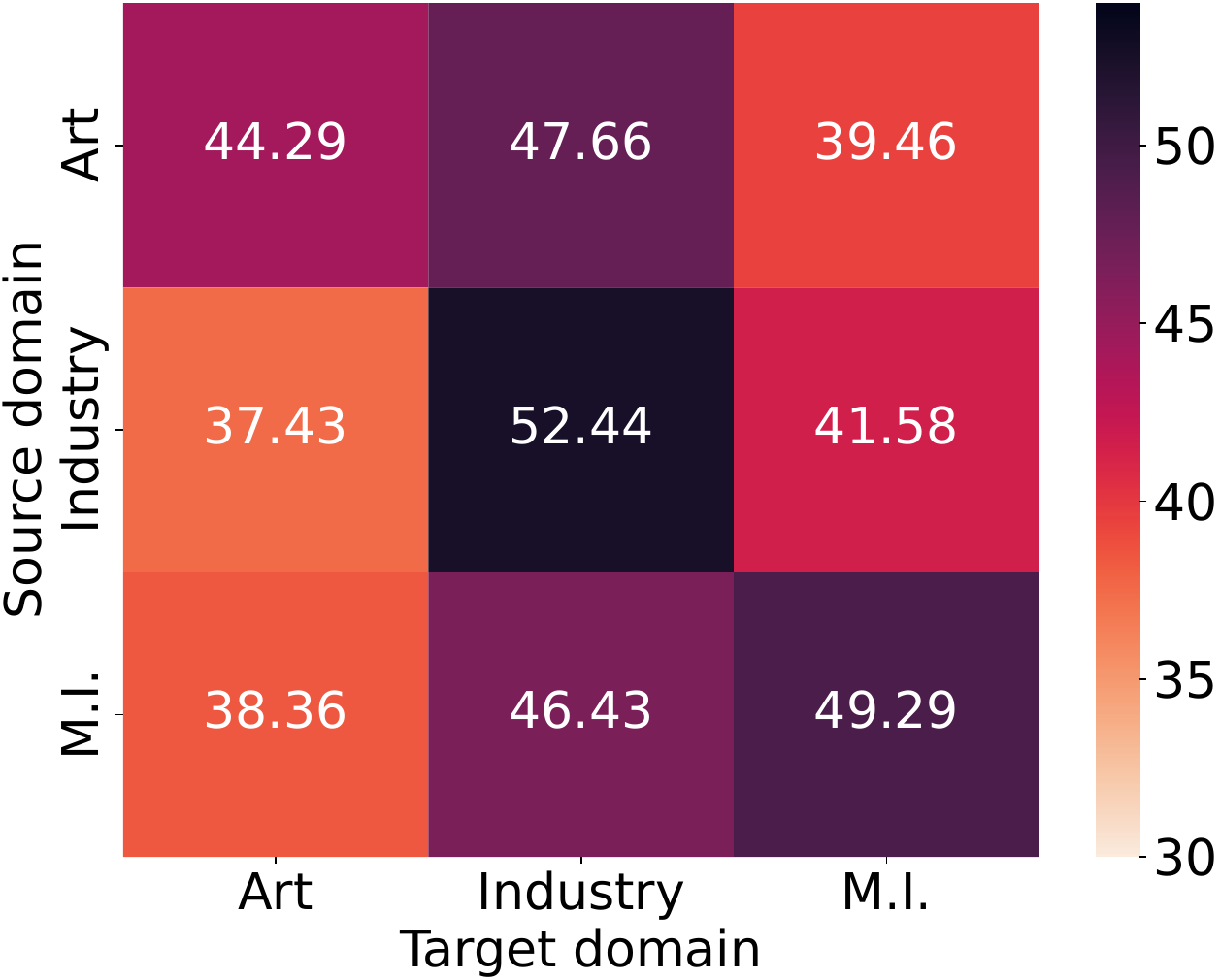}
   }
   \vspace{-2mm}
	\caption{Cross-domain generalization within Amazon. For each method, prompts are selected or tuned on the labeled data of the base classes from a source domain, and tested on the unseen classes from each target domain.}
	\label{fig:domain trainsfer}
	\vspace{-4mm} 
\end{figure}

Firstly, we conduct cross-domain experiments within the Amazon ecosystem. Each Amazon dataset (Art, Industrial and M.I.) represents a different domain. As illustrated in Fig.~\ref{fig:domain trainsfer}, among the three variants, the discrete prompts in \model+d exhibit a suboptimal fit on the source domains, yet they demonstrate a reasonable level of generalization across other domains. Conversely, the learnable static prompts in \model\ fit optimally on the source domains but show minimal generalization ability across other domains. On the other hand, the conditional prompts in \model$^*$ has performed well on both the source and target domains, showing their excellent performance and generalization capabilities.

Secondly, we consider a more challenging cross-domain scenario, where the source and target domains are from completely different ecosystems. 
Specifically, we employ the Amazon datasets as the source domains, and aim to generalize to the target domain on Cora.
Firstly, the left half of Tab.~\ref{table:cross data}, does not involve cross-domain generalization. In other words, both the source domain and the target domain are from the same domain. Specifically, the source domain is the training set of the dataset, and the target domain is its testing set. The results of this part are the few-shot classification results, demonstrating that \model$^*$'s few-shot classification capabilities are on par with G2P2.
Secondly, the right half of Tab.~\ref{table:cross data}, involves cross-domain generalization. This means that the source and target domains are from different datasets. For instance, the source domain might be the training set of the Art, where we fine-tune G2P2’s continuous prompts, or fine-tune both \model$^*$’s continuous prompts and the Meta-net component. We then test on the target domain, which is the testing set of another dataset, Cora. This is an extreme test of the model's generalization ability since the training and testing sets are from entirely different datasets. Indeed, such tests also fall under the category of zero-shot classification. From the results in the right half of Tab.~\ref{table:cross data}, we can see that \model$^*$ performs much better than G2P2 on cross-dataset testing sets, demonstrating the strong generalization capability of \model$^*$'s conditional prompts.

\begin{figure}[tbp]
   \subfigure[Base classes]{
   \centering
   \includegraphics[width=0.3\linewidth]{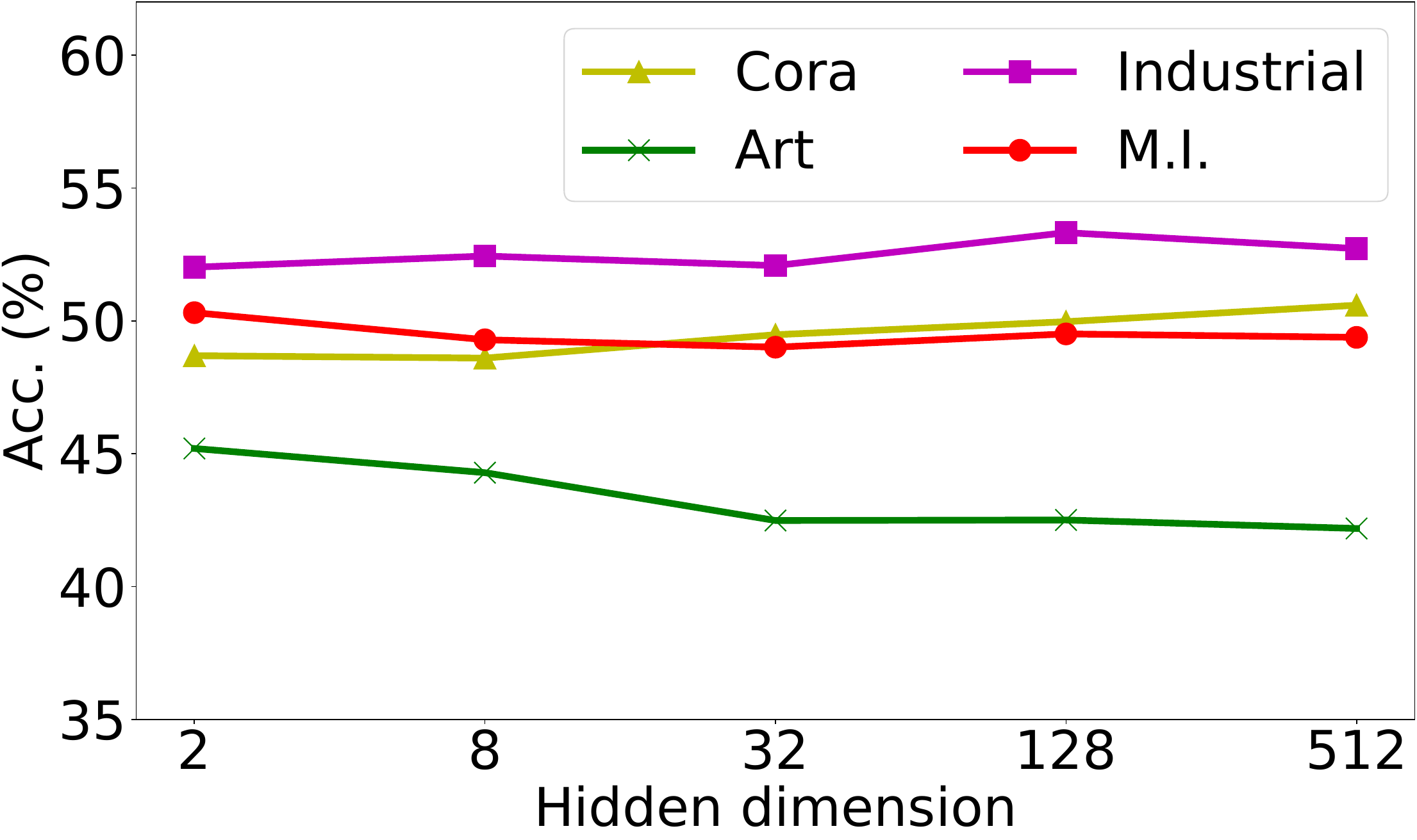}
   }
   \subfigure[Unseen classes]{
   \centering
   \includegraphics[width=0.3\linewidth]{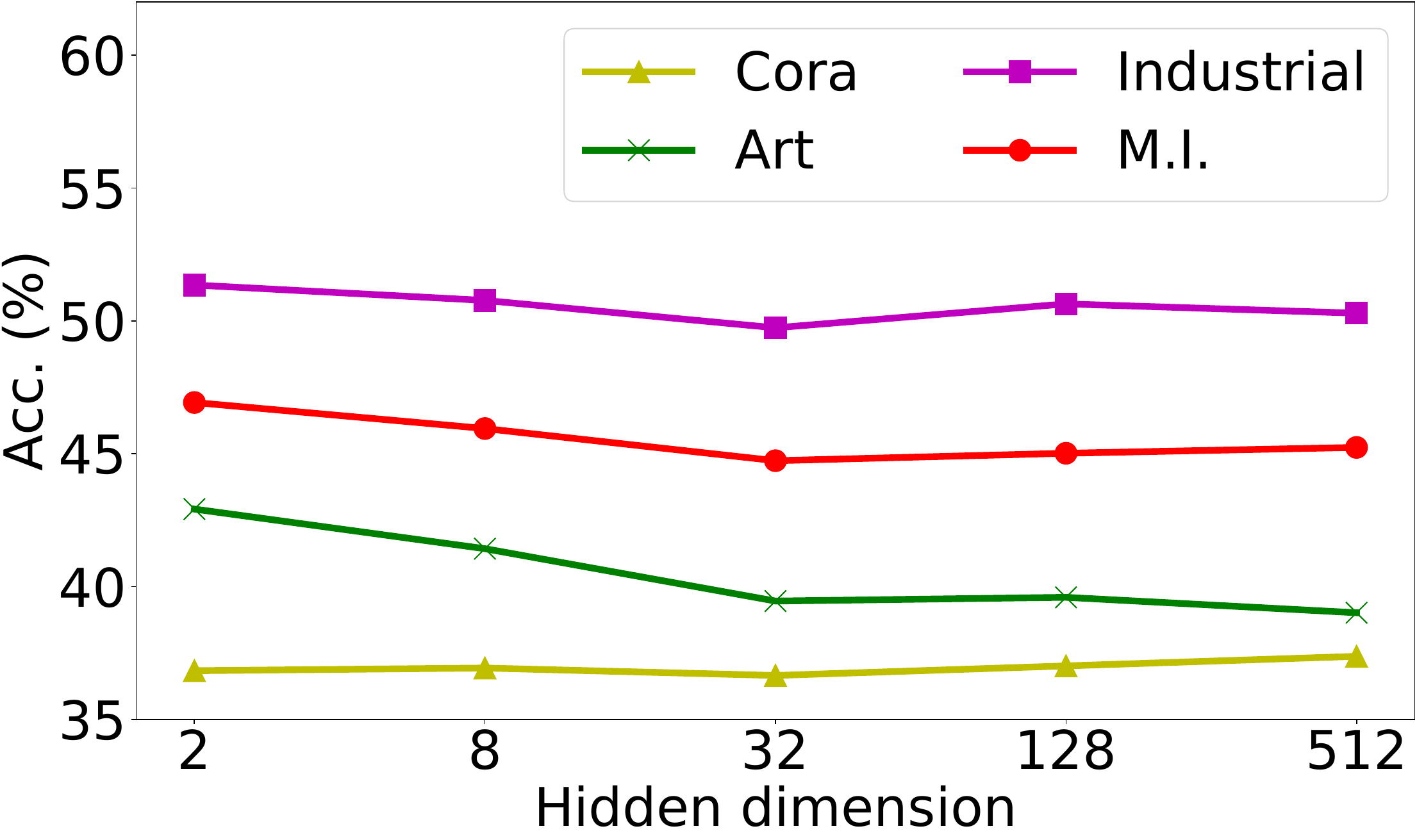}
   }
   \subfigure[Harmonic mean]{
   \centering
   \includegraphics[width=0.3\linewidth]{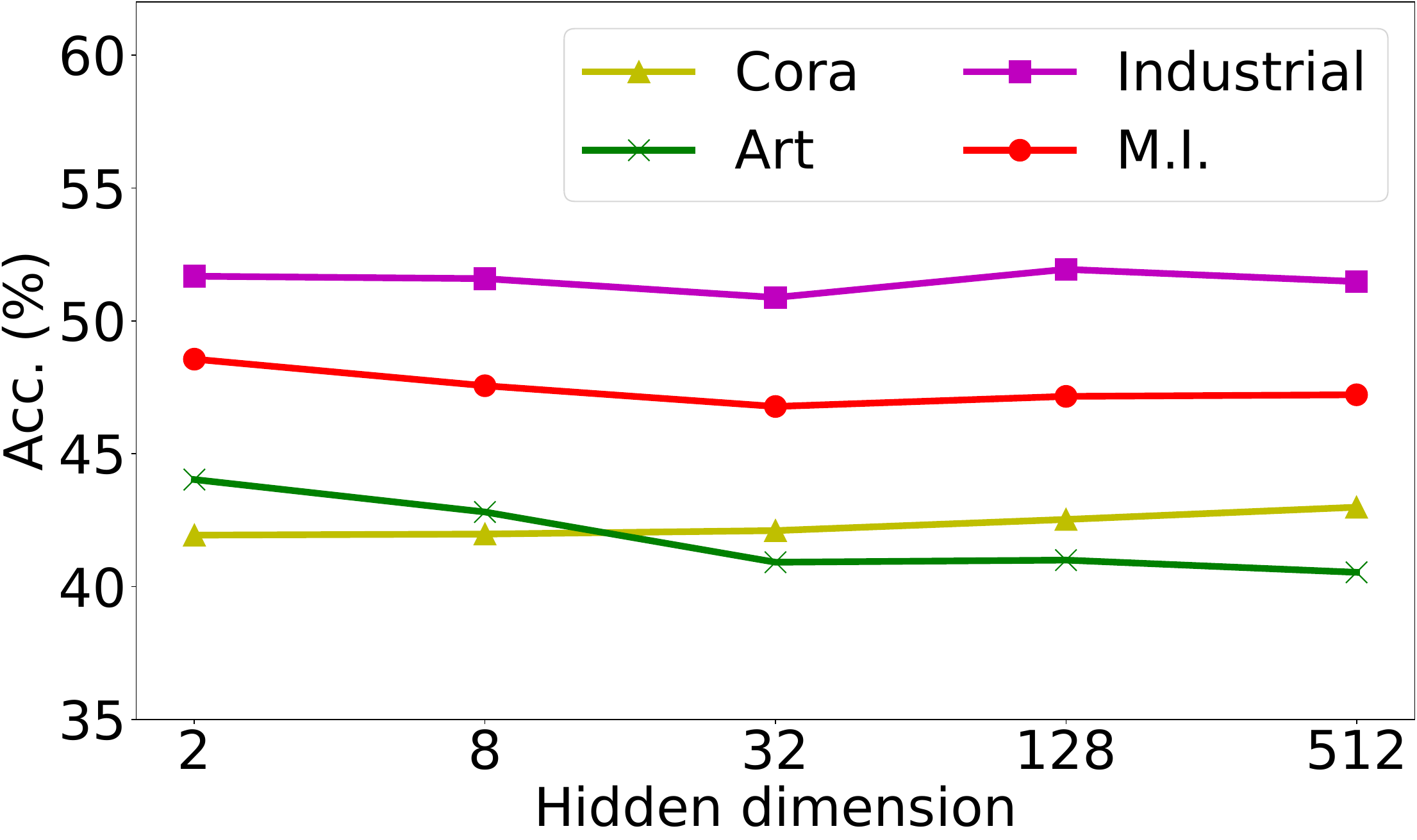}
   }
   \vspace{-2mm}
	\caption{Impact of hidden dimension of Meta-net.}
	\label{fig:param hid_dim}
	\vspace{-2mm} 
\end{figure}

\stitle{Hidden dimension of Meta-net.}
Recall that our proposed \model$^*$ incorporates a Meta-net with a hidden layer. We further study the impact of the hidden dimension of the Meta-net, by varying it between 2 and 512. Observations from Fig.~\ref{fig:param hid_dim} indicate that augmenting the hidden dimension, despite expanding the model capacity, does not yield significantly enhanced performance on both base and unseen classes.
Generally, a larger model with more parameters performs better, but there's an often overlooked assumption: \textbf{sufficient training data}. However, the meta-net of \model$^*$ was trained under a few-shot setting for the base class, \ie, training samples being extremely scarce. Given this significant lack of training data, simply increasing the model size and parameters can lead to overfitting on base classes. Consequently, the performance on unseen classes declines as the hidden dimension of the meta-net increases.
Besides, a larger hidden dimension naturally leads to an increase in computational costs. Consequently, a relatively small value, such as 8, is sufficient yet efficient.

\stitle{Comparing \model$^*$ with a bigger \model~model.}
Given that \model$^*$ incorporates a larger number of parameters than \model, specifically through the addition of the Meta-net, it is important to investigate if the observed improvements by \model$^*$ are merely a consequence of parameter size. Hence, we use more prompt tokens in \model\ so that its number of parameters becomes comparable to that of \model$^*$. Specifically, the number of prompt tokens ($M$) is increased to 16 from 4 in \model, as shown in Tab.~\ref{table:bigger param}.
The results indicate that merely increasing the parameter size does not lead to performance improvement. On the contrary, compared to $M=4$, $M=16$ gives worse performance on both base and unseen classes possibly due to overfitting in low-resource settings. Thus, conditional prompt tuning is a crucial design responsible for the improvements.

\begin{table}[tbp]
    \footnotesize 
	\centering 
 	\addtolength{\tabcolsep}{-2pt}
	\caption{\model$^*$ vs. a bigger \model~model on Art.}
 \vspace{-2mm}
	\label{table:bigger param} 
	\begin{tabular}{l|c|c|c|c}  
		\toprule
		  &Param.~size&Base &Unseen& HM
		 \\\midrule
Model & Parameters & Metric 1 & Metric 2 & Metric 3 \\
\midrule
G2P2 ($M=4$) & 2,048 & \underline{43.47}$\pm$2.59 & \underline{31.97}$\pm$4.69 & \underline{36.71}$\pm$3.73 \\
G2P2 ($M=16$) & \textbf{8,192} & 40.74$\pm$3.31 & 28.67$\pm$2.70 & 33.64$\pm$2.93 \\ 
\model$^*$ ($M=4$) & \underline{7,688} & \textbf{44.29}$\pm$2.30 & \textbf{41.43}$\pm$1.28 & \textbf{42.80}$\pm$1.61 \\

	\bottomrule
	\end{tabular}
\vspace{-2mm}
\end{table}

\section{Conclusion}
In this paper, we studied the problem of low-resource text classification. Given that many text documents are related through an underlying network, we proposed a novel model called Graph-Grounded Pre-training and Prompting (\model). It consists of three graph interaction-based contrastive strategies in pre-training, and a prompting mechanism for the jointly pre-trained graph-text model in downstream classification. We further extended our model to \model$^*$ in order to deal with wider unseen classes. Finally, we conducted extensive experiments and showed the advantages of \model\ and \model$^*$ in low-resource text classification.
We will investigate its application in other text classification scenarios in future work.



\section*{Acknowledgments}
This research / project is supported by the Ministry of Education, Singapore, under its Academic Research Fund Tier 2 (Proposal ID: T2EP20122-0041) and Academic Research Fund Tier 1 grant (22-SIS-SMU-054). Any opinions, findings and conclusions or recommendations expressed in this material are those of the author(s) and do not reflect the views of the Ministry of Education, Singapore.



\bibliographystyle{IEEEtran}
\bibliography{references.bib}

\begin{thebibliography}{10}
\providecommand{\url}[1]{#1}
\csname url@samestyle\endcsname
\providecommand{\newblock}{\relax}
\providecommand{\bibinfo}[2]{#2}
\providecommand{\BIBentrySTDinterwordspacing}{\spaceskip=0pt\relax}
\providecommand{\BIBentryALTinterwordstretchfactor}{4}
\providecommand{\BIBentryALTinterwordspacing}{\spaceskip=\fontdimen2\font plus
\BIBentryALTinterwordstretchfactor\fontdimen3\font minus
  \fontdimen4\font\relax}
\providecommand{\BIBforeignlanguage}[2]{{%
\expandafter\ifx\csname l@#1\endcsname\relax
\typeout{** WARNING: IEEEtran.bst: No hyphenation pattern has been}%
\typeout{** loaded for the language `#1'. Using the pattern for}%
\typeout{** the default language instead.}%
\else
\language=\csname l@#1\endcsname
\fi
#2}}
\providecommand{\BIBdecl}{\relax}
\BIBdecl

\bibitem{mccallum2000automating}
A.~K. McCallum, K.~Nigam, J.~Rennie, and K.~Seymore, ``Automating the
  construction of internet portals with machine learning,'' \emph{Information
  Retrieval}, vol.~3, no.~2, pp. 127--163, 2000.

\bibitem{xu2019open}
H.~Xu, B.~Liu, L.~Shu, and P.~Yu, ``Open-world learning and application to
  product classification,'' in \emph{WWW}, 2019, pp. 3413--3419.

\bibitem{kenton2019bert}
J.~D. M.-W.~C. Kenton and L.~K. Toutanova, ``Bert: Pre-training of deep
  bidirectional transformers for language understanding,'' in \emph{NAACL},
  2019, pp. 4171--4186.

\bibitem{radford2018improving}
A.~Radford, K.~Narasimhan, T.~Salimans, I.~Sutskever \emph{et~al.}, ``Improving
  language understanding by generative pre-training,'' \emph{OpenAI blog},
  2018.

\bibitem{vaswani2017attention}
A.~Vaswani, N.~Shazeer, N.~Parmar, J.~Uszkoreit, L.~Jones, A.~N. Gomez,
  {\L}.~Kaiser, and I.~Polosukhin, ``Attention is all you need,''
  \emph{NeurIPS}, vol.~30, 2017.

\bibitem{brown2020language}
T.~Brown, B.~Mann, N.~Ryder, M.~Subbiah, J.~D. Kaplan, P.~Dhariwal,
  A.~Neelakantan, P.~Shyam, G.~Sastry, A.~Askell \emph{et~al.}, ``Language
  models are few-shot learners,'' \emph{NeurIPS}, vol.~33, pp. 1877--1901,
  2020.

\bibitem{li2021prefix}
X.~L. Li and P.~Liang, ``Prefix-tuning: Optimizing continuous prompts for
  generation,'' in \emph{ACL}, 2021, pp. 4582--4597.

\bibitem{lester2021power}
B.~Lester, R.~Al-Rfou, and N.~Constant, ``The power of scale for
  parameter-efficient prompt tuning,'' in \emph{EMNLP}, 2021, pp. 3045--3059.

\bibitem{liu2021gpt}
X.~Liu, Y.~Zheng, Z.~Du, M.~Ding, Y.~Qian, Z.~Yang, and J.~Tang, ``Gpt
  understands, too,'' \emph{arXiv}, 2021.

\bibitem{wu2020comprehensive}
Z.~Wu, S.~Pan, F.~Chen, G.~Long, C.~Zhang, and S.~Y. Philip, ``A comprehensive
  survey on graph neural networks,'' \emph{TNNLS}, vol.~32, no.~1, pp. 4--24,
  2020.

\bibitem{wu2021self}
L.~Wu, H.~Lin, C.~Tan, Z.~Gao, and S.~Z. Li, ``Self-supervised learning on
  graphs: Contrastive, generative, or predictive,'' \emph{TKDE}, 2021.

\bibitem{velickovic2019deep}
P.~Velickovic, W.~Fedus, W.~L. Hamilton, P.~Li{\`o}, Y.~Bengio, and R.~D.
  Hjelm, ``Deep graph infomax.'' \emph{ICLR}, vol.~2, no.~3, p.~4, 2019.

\bibitem{hu2019strategies}
W.~Hu, B.~Liu, J.~Gomes, M.~Zitnik, P.~Liang, V.~Pande, and J.~Leskovec,
  ``Strategies for pre-training graph neural networks,'' in \emph{ICLR}, 2019.

\bibitem{hu2020gpt}
Z.~Hu, Y.~Dong, K.~Wang, K.-W. Chang, and Y.~Sun, ``Gpt-gnn: Generative
  pre-training of graph neural networks,'' in \emph{KDD}, 2020, pp. 1857--1867.

\bibitem{yang2016revisiting}
Z.~Yang, W.~Cohen, and R.~Salakhudinov, ``Revisiting semi-supervised learning
  with graph embeddings,'' in \emph{ICML}, 2016, pp. 40--48.

\bibitem{mikolov2013efficient}
T.~Mikolov, K.~Chen, G.~Corrado, and J.~Dean, ``Efficient estimation of word
  representations in vector space,'' \emph{arXiv}, 2013.

\bibitem{zhou2022conditional}
K.~Zhou, J.~Yang, C.~C. Loy, and Z.~Liu, ``Conditional prompt learning for
  vision-language models,'' in \emph{CVPR}, 2022, pp. 16\,816--16\,825.

\bibitem{wen2023augmenting}
Z.~Wen and Y.~Fang, ``Augmenting low-resource text classification with
  graph-grounded pre-training and prompting,'' \emph{arXiv}, 2023.

\bibitem{KipfW17}
T.~N. Kipf and M.~Welling, ``Semi-supervised classification with graph
  convolutional networks,'' in \emph{ICLR}.\hskip 1em plus 0.5em minus
  0.4em\relax OpenReview.net, 2017.

\bibitem{velivckovic2018graph}
P.~Veli{\v{c}}kovi{\'c}, G.~Cucurull, A.~Casanova, A.~Romero, P.~Li{\`o}, and
  Y.~Bengio, ``Graph attention networks,'' in \emph{ICLR}, 2018.

\bibitem{XuHLJ19}
K.~Xu, W.~Hu, J.~Leskovec, and S.~Jegelka, ``How powerful are graph neural
  networks?'' in \emph{ICLR}.\hskip 1em plus 0.5em minus 0.4em\relax
  OpenReview.net, 2019.

\bibitem{lu2021learning}
Y.~Lu, X.~Jiang, Y.~Fang, and C.~Shi, ``Learning to pre-train graph neural
  networks,'' in \emph{AAAI}, vol.~35, no.~5, 2021, pp. 4276--4284.

\bibitem{cao2021dekr}
X.~Cao, Y.~Shi, H.~Yu, J.~Wang, X.~Wang, Z.~Yan, and Z.~Chen, ``Dekr:
  description enhanced knowledge graph for machine learning method
  recommendation,'' in \emph{SIGIR}, 2021, pp. 203--212.

\bibitem{linmei2019heterogeneous}
H.~Linmei, T.~Yang, C.~Shi, H.~Ji, and X.~Li, ``Heterogeneous graph attention
  networks for semi-supervised short text classification,'' in \emph{EMNLP},
  2019, pp. 4821--4830.

\bibitem{liu2021mm}
Y.~Liu, S.~Yang, C.~Lei, G.~Wang, H.~Tang, J.~Zhang, A.~Sun, and C.~Miao,
  ``Pre-training graph transformer with multimodal side information for
  recommendation,'' in \emph{ACMMM}, 2021b, pp. 2853--2861.

\bibitem{zhao2023learning}
J.~Zhao, M.~Qu, C.~Li, H.~Yan, Q.~Liu, R.~Li, X.~Xie, and J.~Tang, ``Learning
  on large-scale text-attributed graphs via variational inference,'' in
  \emph{ICLR}, 2023.

\bibitem{han2021pre}
X.~Han, Z.~Zhang, N.~Ding, Y.~Gu, X.~Liu, Y.~Huo, J.~Qiu, Y.~Yao, A.~Zhang,
  L.~Zhang \emph{et~al.}, ``Pre-trained models: Past, present and future,''
  \emph{AI Open}, vol.~2, pp. 225--250, 2021.

\bibitem{yang2019xlnet}
Z.~Yang, Z.~Dai, Y.~Yang, J.~Carbonell, R.~R. Salakhutdinov, and Q.~V. Le,
  ``Xlnet: Generalized autoregressive pretraining for language understanding,''
  \emph{NeurIPS}, vol.~32, 2019.

\bibitem{liu2019roberta}
Y.~Liu, M.~Ott, N.~Goyal, J.~Du, M.~Joshi, D.~Chen, O.~Levy, M.~Lewis,
  L.~Zettlemoyer, and V.~Stoyanov, ``{RoBERTa}: A robustly optimized bert
  pretraining approach,'' \emph{arXiv}, 2019.

\bibitem{raffel2020exploring}
C.~Raffel, N.~Shazeer, A.~Roberts, K.~Lee, S.~Narang, M.~Matena, Y.~Zhou,
  W.~Li, P.~J. Liu \emph{et~al.}, ``Exploring the limits of transfer learning
  with a unified text-to-text transformer.'' \emph{J. Mach. Learn. Res.},
  vol.~21, no. 140, pp. 1--67, 2020.

\bibitem{liu2021pre}
P.~Liu, W.~Yuan, J.~Fu, Z.~Jiang, H.~Hayashi, and G.~Neubig, ``Pre-train,
  prompt, and predict: A systematic survey of prompting methods in natural
  language processing,'' \emph{arXiv}, 2021.

\bibitem{shin2020autoprompt}
T.~Shin, Y.~Razeghi, R.~L. Logan~IV, E.~Wallace, and S.~Singh, ``{AutoPrompt}:
  Eliciting knowledge from language models with automatically generated
  prompts,'' in \emph{EMNLP}, 2020, pp. 4222--4235.

\bibitem{schick2021exploiting}
T.~Schick and H.~Sch{\"u}tze, ``Exploiting cloze-questions for few-shot text
  classification and natural language inference,'' in \emph{EACL}, 2021, pp.
  255--269.

\bibitem{gao2021making}
T.~Gao, A.~Fisch, and D.~Chen, ``Making pre-trained language models better
  few-shot learners,'' in \emph{ACL}, 2021, pp. 3816--3830.

\bibitem{qin2021learning}
G.~Qin and J.~Eisner, ``Learning how to ask: Querying lms with mixtures of soft
  prompts,'' in \emph{NAACL}, 2021, pp. 5203--5212.

\bibitem{zhong2021factual}
Z.~Zhong, D.~Friedman, and D.~Chen, ``Factual probing is [{MASK}]: Learning vs.
  learning to recall,'' in \emph{NAACL}, 2021, pp. 5017--5033.

\bibitem{hu2022knowledgeable}
S.~Hu, N.~Ding, H.~Wang, Z.~Liu, J.~Wang, J.~Li, W.~Wu, and M.~Sun,
  ``Knowledgeable prompt-tuning: Incorporating knowledge into prompt verbalizer
  for text classification,'' in \emph{ACL}, 2022, pp. 2225--2240.

\bibitem{min2022noisy}
S.~Min, M.~Lewis, H.~Hajishirzi, and L.~Zettlemoyer, ``Noisy channel language
  model prompting for few-shot text classification,'' in \emph{ACL}, 2022, pp.
  5316--5330.

\bibitem{sun2021nsp}
Y.~Sun, Y.~Zheng, C.~Hao, and H.~Qiu, ``{NSP-BERT}: A prompt-based zero-shot
  learner through an original pre-training task--next sentence prediction,''
  \emph{arXiv}, 2021.

\bibitem{zhang2021aspect}
W.~Zhang, Y.~Deng, X.~Li, Y.~Yuan, L.~Bing, and W.~Lam, ``Aspect sentiment quad
  prediction as paraphrase generation,'' in \emph{EMNLP}, 2021, pp. 9209--9219.

\bibitem{han2021ptr}
X.~Han, W.~Zhao, N.~Ding, Z.~Liu, and M.~Sun, ``{PTR}: Prompt tuning with rules
  for text classification,'' \emph{arXiv}, 2021.

\bibitem{tan2022msp}
Z.~Tan, X.~Zhang, S.~Wang, and Y.~Liu, ``{MSP}: Multi-stage prompting for
  making pre-trained language models better translators,'' in \emph{ACL}, 2022,
  pp. 6131--6142.

\bibitem{chen2022knowprompt}
X.~Chen, N.~Zhang, X.~Xie, S.~Deng, Y.~Yao, C.~Tan, F.~Huang, L.~Si, and
  H.~Chen, ``{KnowPrompt}: Knowledge-aware prompt-tuning with synergistic
  optimization for relation extraction,'' in \emph{WWW}, 2022, pp. 2778--2788.

\bibitem{sainz2021label}
O.~Sainz, O.~L. de~Lacalle, G.~Labaka, A.~Barrena, and E.~Agirre, ``Label
  verbalization and entailment for effective zero and few-shot relation
  extraction,'' in \emph{EMNLP}, 2021, pp. 1199--1212.

\bibitem{sun2022gppt}
M.~Sun, K.~Zhou, X.~He, Y.~Wang, and X.~Wang, ``{GPPT}: Graph pre-training and
  prompt tuning to generalize graph neural networks,'' in \emph{KDD}, 2022, pp.
  1717--1727.

\bibitem{tan2023virtual}
Z.~Tan, R.~Guo, K.~Ding, and H.~Liu, ``Virtual node tuning for few-shot node
  classification,'' \emph{arXiv}, 2023.

\bibitem{fang2022prompt}
T.~Fang, Y.~Zhang, Y.~Yang, and C.~Wang, ``Prompt tuning for graph neural
  networks,'' \emph{arXiv}, 2022.

\bibitem{liu2023graphprompt}
Z.~Liu, X.~Yu, Y.~Fang, and X.~Zhang, ``{GraphPrompt}: Unifying pre-training
  and downstream tasks for graph neural networks,'' in \emph{WWW}, 2023, pp.
  417--428.

\bibitem{MiyatoDG17}
T.~Miyato, A.~M. Dai, and I.~J. Goodfellow, ``Adversarial training methods for
  semi-supervised text classification,'' in \emph{ICLR}.\hskip 1em plus 0.5em
  minus 0.4em\relax OpenReview.net, 2017.

\bibitem{xie2020unsupervised}
Q.~Xie, Z.~Dai, E.~Hovy, T.~Luong, and Q.~Le, ``Unsupervised data augmentation
  for consistency training,'' \emph{NeurIPS}, vol.~33, pp. 6256--6268, 2020.

\bibitem{chen2020mixtext}
J.~Chen, Z.~Yang, and D.~Yang, ``{MixText}: Linguistically-informed
  interpolation of hidden space for semi-supervised text classification,'' in
  \emph{ACL}, 2020, pp. 2147--2157.

\bibitem{finn2017model}
C.~Finn, P.~Abbeel, and S.~Levine, ``Model-agnostic meta-learning for fast
  adaptation of deep networks,'' in \emph{ICML}, 2017, pp. 1126--1135.

\bibitem{yu2018diverse}
M.~Yu, X.~Guo, J.~Yi, S.~Chang, S.~P. Y. C.~G. Tesauro, H.~W.~B. Zhou, and
  A.~Foundations-Learning, ``Diverse few-shot text classification with multiple
  metrics,'' in \emph{NAACL}, 2018, pp. 1206--1215.

\bibitem{han2018fewrel}
X.~Han, H.~Zhu, P.~Yu, Z.~Wang, Y.~Yao, Z.~Liu, and M.~Sun, ``{FewRel}: A
  large-scale supervised few-shot relation classification dataset with
  state-of-the-art evaluation,'' in \emph{EMNLP}, 2018, pp. 4803--4809.

\bibitem{bansal2020self}
T.~Bansal, R.~Jha, T.~Munkhdalai, and A.~McCallum, ``Self-supervised
  meta-learning for few-shot natural language classification tasks,'' in
  \emph{EMNLP}, 2020, pp. 522--534.

\bibitem{bao2020few}
Y.~Bao, M.~Wu, S.~Chang, and R.~Barzilay, ``Few-shot text classification with
  distributional signatures,'' in \emph{ICLR}, 2020.

\bibitem{zhou2019meta}
F.~Zhou, C.~Cao, K.~Zhang, G.~Trajcevski, T.~Zhong, and J.~Geng, ``{Meta-GNN}:
  On few-shot node classification in graph meta-learning,'' in \emph{CIKM},
  2019, pp. 2357--2360.

\bibitem{wang2020graph}
N.~Wang, M.~Luo, K.~Ding, L.~Zhang, J.~Li, and Q.~Zheng, ``Graph few-shot
  learning with attribute matching,'' in \emph{CIKM}, 2020, pp. 1545--1554.

\bibitem{wen2021meta}
Z.~Wen, Y.~Fang, and Z.~Liu, ``Meta-inductive node classification across
  graphs,'' in \emph{SIGIR}, 2021, pp. 1219--1228.

\bibitem{wen2023generalizing}
Z.~Wen, ``Generalizing graph neural network across graphs and time,'' in
  \emph{WSDM}, 2023, pp. 1214--1215.

\bibitem{radford2021learning}
A.~Radford, J.~W. Kim, C.~Hallacy, A.~Ramesh, G.~Goh, S.~Agarwal, G.~Sastry,
  A.~Askell, P.~Mishkin, J.~Clark \emph{et~al.}, ``Learning transferable visual
  models from natural language supervision,'' in \emph{ICML}, 2021, pp.
  8748--8763.

\bibitem{zhou2022learning}
K.~Zhou, J.~Yang, C.~C. Loy, and Z.~Liu, ``Learning to prompt for
  vision-language models,'' \emph{IJCV}, vol. 130, no.~9, pp. 2337--2348, 2022.

\bibitem{wang2021zero}
Z.~Wang, J.~Wang, Y.~Guo, and Z.~Gong, ``Zero-shot node classification with
  decomposed graph prototype network,'' in \emph{KDD}, 2021, pp. 1769--1779.

\bibitem{liu2018content}
J.~Liu, Z.~He, L.~Wei, and Y.~Huang, ``Content to node: Self-translation
  network embedding,'' in \emph{KDD}, 2018, pp. 1794--1802.

\bibitem{sohn2016improved}
K.~Sohn, ``Improved deep metric learning with multi-class n-pair loss
  objective,'' \emph{NeurIPS}, vol.~29, 2016.

\bibitem{zhang2020contrastive}
Y.~Zhang, H.~Jiang, Y.~Miura, C.~D. Manning, and C.~P. Langlotz, ``Contrastive
  learning of medical visual representations from paired images and text,''
  \emph{arXiv}, 2020.

\bibitem{wang2018joint}
G.~Wang, C.~Li, W.~Wang, Y.~Zhang, D.~Shen, X.~Zhang, R.~Henao, and L.~Carin,
  ``Joint embedding of words and labels for text classification,'' in
  \emph{ACL}, 2018, pp. 2321--2331.

\bibitem{liu-etal-2022-p}
X.~Liu, K.~Ji, Y.~Fu, W.~Tam, Z.~Du, Z.~Yang, and J.~Tang, ``{P}-tuning: Prompt
  tuning can be comparable to fine-tuning across scales and tasks,'' in
  \emph{ACL}, 2022, pp. 61--68.

\bibitem{ha2017hypernetworks}
D.~Ha, A.~M. Dai, and Q.~V. Le, ``Hypernetworks,'' in \emph{ICLR}, 2017.

\bibitem{wu2018reducing}
Y.~Wu and T.~Lee, ``Reducing model complexity for dnn based large-scale audio
  classification,'' in \emph{ICASSP}, 2018, pp. 331--335.

\bibitem{hu2021lora}
E.~J. Hu, Y.~Shen, P.~Wallis, Z.~Allen-Zhu, Y.~Li, S.~Wang, L.~Wang, and
  W.~Chen, ``{LoRA}: Low-rank adaptation of large language models,''
  \emph{arXiv}, 2021.

\bibitem{ni2019justifying}
J.~Ni, J.~Li, and J.~McAuley, ``Justifying recommendations using
  distantly-labeled reviews and fine-grained aspects,'' in \emph{EMNLP}, 2019,
  pp. 188--197.

\bibitem{hamilton2017inductive}
W.~Hamilton, Z.~Ying, and J.~Leskovec, ``Inductive representation learning on
  large graphs,'' in \emph{NeurIPS}, 2017.

\bibitem{yao2019graph}
L.~Yao, C.~Mao, and Y.~Luo, ``Graph convolutional networks for text
  classification,'' in \emph{AAAI}, vol.~33, no.~01, 2019, pp. 7370--7377.

\bibitem{tian2020rethinking}
Y.~Tian, Y.~Wang, D.~Krishnan, J.~B. Tenenbaum, and P.~Isola, ``Rethinking
  few-shot image classification: a good embedding is all you need?'' in
  \emph{ECCV}, 2020, pp. 266--282.

\bibitem{bai2023qwen}
J.~Bai, S.~Bai, Y.~Chu, Z.~Cui, K.~Dang, X.~Deng, Y.~Fan, W.~Ge, Y.~Han,
  F.~Huang \emph{et~al.}, ``Qwen technical report,'' \emph{arXiv}, 2023.

\bibitem{muennighoff2022crosslingual}
N.~Muennighoff, T.~Wang, L.~Sutawika, A.~Roberts, S.~Biderman, T.~L. Scao,
  M.~S. Bari, S.~Shen, Z.-X. Yong, H.~Schoelkopf \emph{et~al.}, ``Crosslingual
  generalization through multitask finetuning,'' \emph{arXiv}, 2022.

\bibitem{llama3}
``Llama 3 model card,''
  \emph{https://llama.meta.com/docs/model-cards-and-prompt-formats/meta-llama-3/Llama
  3 Model Card}, 2024.

\bibitem{baichuan13b}
``A 13b large language model developed by baichuan intelligent technology.''
  \emph{https://github.com/baichuan-inc/Baichuan-13B}, 2024.

\bibitem{zheng2024judging}
L.~Zheng, W.-L. Chiang, Y.~Sheng, S.~Zhuang, Z.~Wu, Y.~Zhuang, Z.~Lin, Z.~Li,
  D.~Li, E.~Xing \emph{et~al.}, ``Judging {LLM}-as-a-judge with {MT}-bench and
  chatbot arena,'' \emph{NeuraIPS}, vol.~36, 2024.

\bibitem{achiam2023gpt}
J.~Achiam, S.~Adler, S.~Agarwal, L.~Ahmad, I.~Akkaya, F.~L. Aleman, D.~Almeida,
  J.~Altenschmidt, S.~Altman, S.~Anadkat \emph{et~al.}, ``{GPT-4} technical
  report,'' \emph{arXiv}, 2023.

\bibitem{bard}
``An overview of bard: an early experiment with generative ai,''
  \emph{https://ai.google/static/documents/google-about-bard.pdf}, 2024.

\bibitem{Alpaca}
``Alpaca: A strong, replicable instruction-following model,''
  \emph{https://crfm.stanford.edu/2023/03/13/alpaca.html}, 2024.

\bibitem{sennrich2016neural}
R.~Sennrich, B.~Haddow, and A.~Birch, ``Neural machine translation of rare
  words with subword units,'' in \emph{ACL}, 2016, pp. 1715--1725.

\bibitem{perozzi2014deepwalk}
B.~Perozzi, R.~Al-Rfou, and S.~Skiena, ``{DeepWalk}: Online learning of social
  representations,'' in \emph{KDD}, 2014, pp. 701--710.

\bibitem{wolf2020transformers}
T.~Wolf, L.~Debut, V.~Sanh, J.~Chaumond, C.~Delangue, A.~Moi, P.~Cistac,
  T.~Rault, R.~Louf, M.~Funtowicz \emph{et~al.}, ``Transformers:
  State-of-the-art natural language processing,'' in \emph{EMNLP}, 2020, pp.
  38--45.

\bibitem{xian2017zero}
Y.~Xian, B.~Schiele, and Z.~Akata, ``Zero-shot learning-the good, the bad and
  the ugly,'' in \emph{CVPR}, 2017, pp. 4582--4591.

\bibitem{parisi2019continual}
G.~I. Parisi, R.~Kemker, J.~L. Part, C.~Kanan, and S.~Wermter, ``Continual
  lifelong learning with neural networks: A review,'' \emph{Neural networks},
  vol. 113, pp. 54--71, 2019.

\end{thebibliography}


 



\vspace{-10mm}
\begin{IEEEbiographynophoto}{Zhihao Wen}
received his Ph.D. degree in Computer Science from the School of Computing and Information Systems, Singapore Management University. His research interests include graph-based machine learning and data mining, as well as their applications for the Web and social media.
\end{IEEEbiographynophoto}

\vspace{-10mm}
\begin{IEEEbiographynophoto}{Yuan Fang}
received his Ph.D. degree in Computer Science from University of Illinois at Urbana Champaign in 2014. He is currently an Assistant Professor in the School of Computing and Information Systems, Singapore Management University. His research focuses on graph-based machine learning and data mining, as well as their applications for the Web and social media.
\end{IEEEbiographynophoto}

\vfill

\end{document}